\documentclass[sigconf]{acmart}

\AtBeginDocument{%
  }

\renewcommand\footnotetextcopyrightpermission[1]{} 
\acmConference{}
\acmBooktitle{}

\acmSubmissionID{}

\usepackage[export]{adjustbox} 
\usepackage{tikz}
\usepackage{listings}
\usepackage[toc,page]{appendix}
\newcommand{\filledcircle}[1]{\tikz[baseline=(char.base)]{%
    \node[shape=circle,fill=black,text=white,inner sep=0.5pt,minimum size=3.5mm] (char) {\tiny\bfseries#1};}}

\newcommand{\methodname}[0]{RELATE-Sim}
\renewcommand\footnotetextcopyrightpermission[1]{} 
\makeatletter
\def\@authorfont{\normalfont\normalsize}
\def\@affiliationfont{\normalfont\normalsize}
\makeatother

\begin{document}

\title[RELATE-Sim]{RELATE-Sim: Leveraging Turning Point Theory and LLM Agents to Predict and Understand Long-Term Relationship Dynamics through Interactive Narrative Simulations}

\author{\textbf{Matthew Yue, Zhikun Xu, Vivek Gupta, Thao Ha, Liesel Sharabi, Ben Zhou}}
\affiliation{%
\institution{Arizona State University}
\city{Tempe}
\state{Arizona}
\country{United States}
}

\begin{abstract}
Most dating technologies optimize for getting together, not staying together. We present RELATE-Sim, a theory-grounded simulator that models how couples behave at consequential turning points—exclusivity talks, conflict-and-repair episodes, relocations—rather than static traits. Two persona-aligned LLM agents (one per partner) interact under a centralized Scene Master that frames each turning point as a compact set of realistic options, advances the narrative, and infers interpretable state changes and an auditable commitment estimate after each scene. On a longitudinal dataset of 71 couples with two-year follow-ups, simulation-aware predictions outperform a personas-only baseline while surfacing actionable markers (e.g., repair attempts acknowledged, clarity shifts) that explain why trajectories diverge. RELATE-Sim pushes the relationship research’s focus from matchmaking to maintenance, providing a transparent, extensible platform for understanding and forecasting long-term relationship dynamics.
\end{abstract}



\begin{teaserfigure}
  \centering
  \begin{adjustbox}{max size={\textwidth}{.95\textheight},center}
    \includegraphics[clip,trim=6mm 15mm 5mm 21mm]{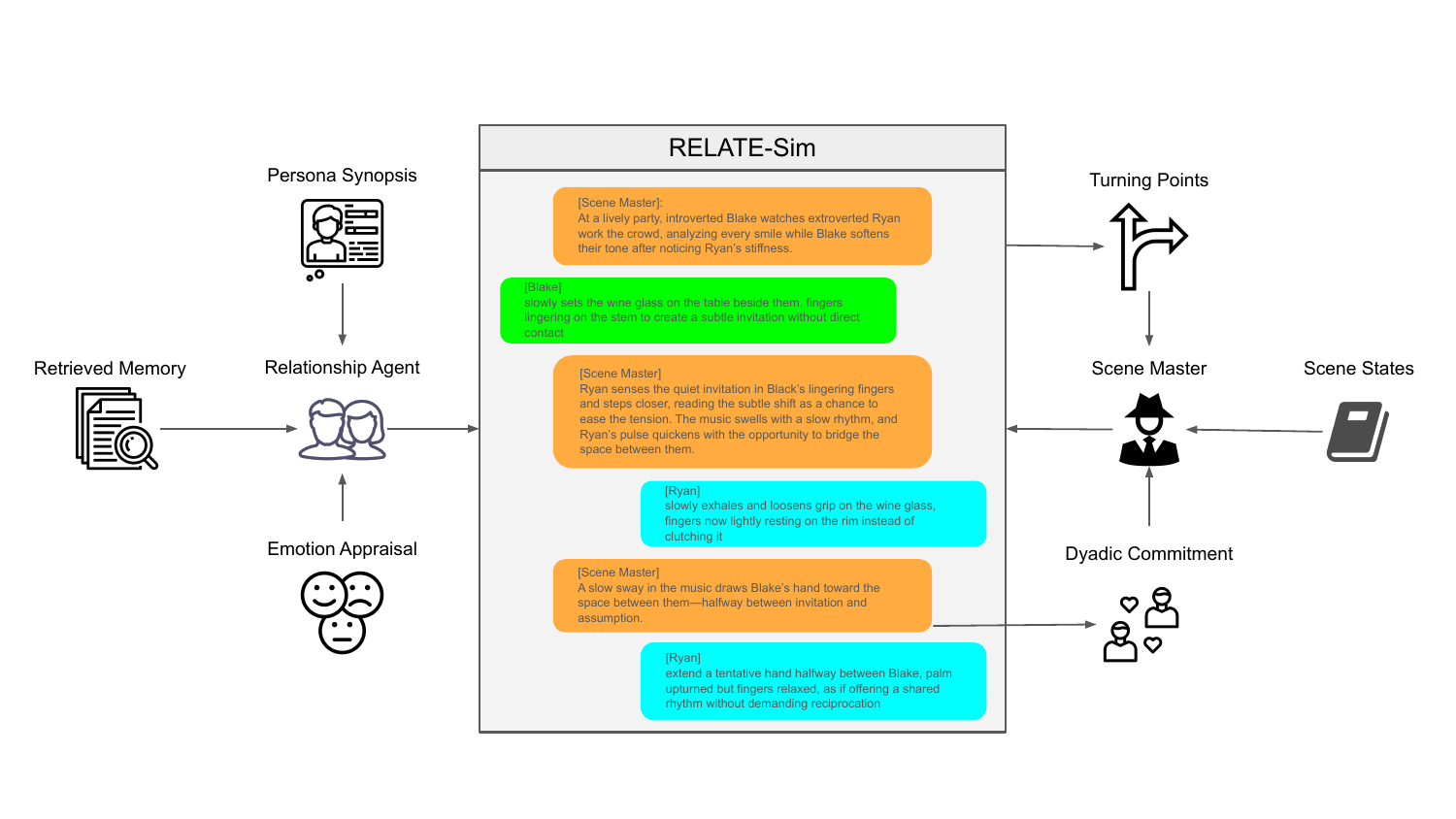} 
  \end{adjustbox}
  \caption{\textbf{The RELATE-Sim framework overview and example outputs. RELATE-Sim stages realistic ``turning-point'' scenes (exclusivity, conflict, life transitions) between two persona-aligned LLM agents and then summarizes the interaction in interpretable states (e.g., conflict, repair, clarity). From intake to scene selection to after-action feedback, the system yields actionable signals for couples and features that forecast future commitments, providing a closer look at how rehearsal can anticipate real-world dynamics.}}
  \label{fig:teaser}
\end{teaserfigure}


\maketitle
\pagestyle{plain}

\section{Introduction}
Digital tools have transformed how people meet, yet most systems still optimize for the start of a relationship—matching on profiles, preferences, or first-date chemistry—rather than for what actually sustains commitment over months and years \cite{doi:10.1177/1529100612436522,Sharabi2022FindingLO,whatpredictsfirstdate}. A large body of social-science and HCI research shows that long-term outcomes are shaped less by static traits and more by interactional processes: how partners navigate conflict, set boundaries, weather stressors, and renegotiate commitments at consequential moments \cite{arriaga2006relationship,gottman2014predicts,feelingloved,emotionregulation}. Today’s LLM agents and social simulations \cite{zhou2024sotopia, DBLP:conf/uist/ParkOCMLB23} hint at a path forward, but existing work rarely models the dyadic mechanisms that drive commitment change or connects simulated behavior to externally verifiable trajectories.

In this paper, we argue that understanding (and eventually supporting) long-term romantic relationships requires models that (i) are grounded in each partner’s persona and lived constraints, (ii) surface turning points—pivotal events that recalibrate norms and commitment—and (iii) make those events actionable through interpretable decisions and outcomes. Rather than asking “are these two compatible on paper?”, we ask ``how do these two behave when it matters, and what does that imply for the path ahead?''

RELATE-Sim addresses this gap with a theory-grounded, interactive simulation of dyadic dynamics. Our system instantiates each partner as a persona-aligned LLM agent and places the pair into a sequence of consequential, turning-point scenarios (e.g., exclusivity talks, conflict-and-repair episodes, relocations, major life transitions). A centralized turning-point–driven, text-game “Scene Master” frames each scenario, curates a small set of realistic decision options, advances the narrative, and infers interpretable state changes (e.g., clarity, repair, alternatives, constraints). Across scenes, the simulation yields auditable traces of behavior and a theory-linked estimate of commitment that can be aggregated to predict longer-term outcomes.

This framing brings three advantages. First, it aligns with decades of relationship science on commitment, interaction patterns, and the diagnostic value of turning points. Second, it makes the simulation’s internal logic legible to both researchers and participants: options are concrete, states are named, and inferences are evidence-linked. Third, it shifts relationship technology from matchmaking toward development and maintenance—a design space that HCI is well-positioned to explore.

This paper makes the following contributions:
\begin{enumerate}
    \item \textbf{Persona-based Simulation Framework.} We present a pipeline that synthesizes multi-source inputs into concise, evidence-linked partner personas and a small rule-based “playbook.” These personas ground agent behavior and preserve actor–partner divergences that matter for dyadic reasoning.
    \item \textbf{Turning-point–driven Text-game Manager.} We introduce a Scene Master that operationalizes turning-point theory as an interactive narrative: it seeds realistic scenarios, constrains choices to 3–4 auditable options, and tracks interpretable relationship states across scenes. This design provides safe “rehearsal” for high-stakes moments while enabling consistent analysis.
    \item \textbf{Theory-grounded Commitment Inference.} We link observable behaviors to commitment via explicit rubrics derived from established frameworks (e.g., investments, alternatives, conflict-repair), yielding per-scene scores and rationales that can be aggregated into trajectory predictions.
    \item \textbf{Empirical Evaluation on Longitudinal Couples.} Using a dataset with baseline measures and two-year follow-ups, we show that simulation-aware predictions outperform a personas-only baseline for forecasting relationship outcomes. Beyond accuracy, the simulation widens between-group separation and surfaces actionable markers that explain why trajectories differ.
\end{enumerate}

Together, these contributions demonstrate a practical, interpretable path to modeling long-term relationship dynamics with LLM agents. Conceptually, we recast “compatibility” as competence at consequential moments; methodologically, we combine persona grounding with a turning-point narrative engine to elicit those moments; and empirically, we connect simulated behavior to externally verified change in relationship status. On $N=101$ couples with two-year follow-ups, the simulation-aware predictor outperformed a personas-only baseline (64.4\% vs. 48.5\% accuracy); on a separate $N=71$ evaluation, simulated results increased between-group (improved vs deteriorated relationship status couples) separation 3x. The scene traces revealed actionable markers—repair attempts acknowledged, clarity gains after turning-point talks, and alternative salience—that align with commitment theory and make the model’s inferences auditable.

RELATE-Sim is not a matchmaking algorithm nor a therapeutic intervention. It is a research platform for understanding and forecasting relationship dynamics and for exploring design patterns that might one day support couples before high-stakes events occur. Our present evaluation focuses on text-only simulations seeded from baseline data and does not involve real-time user participation. We discuss extensions—human-in-the-loop decision points, richer third-party and stressor models, and prospective data collection—that open opportunities for participatory, ethical, and culturally inclusive HCI research on enduring relationships. 

In sum, this work shifts attention from first-date fit to long-horizon functioning. By combining persona-grounded agents with a turning-point–driven text-game manager, we provide a rigorous, interpretable, and extensible way to study how couples behave when it matters—and what those behaviors imply for commitment over time.

\section{Related Works}

\subsection{Generative Agents \& Social Simulations}
LLM use has progressed from single-/few-turn assistants to RAG-grounded enterprise chat and tool-augmented agents that orchestrate workflows; within this arc, generative-agent research shows how memory and reflection produce believable social routines and collective events (e.g., Park et al.’s sandbox simulation \cite{DBLP:conf/uist/ParkOCMLB23}). Persona-alignment studies find that assigned personalities persist across tasks via Big Five markers, psychology-guided life histories, and analyses of role-play validity under complexity \cite{shao-etal-2023-character, jiang-etal-2024-personallm, wang-etal-2024-rolellm}. Multi-agent work probes collaboration and social dynamics, from trait-diverse societies and strategy comparisons \cite{city35211} to inception-prompted role cooperation (e.g., CAMEL \cite{li2023camel}) and open-ended social benchmarks like SOTOPIA \cite{zhou2024sotopia}. Role-playing frameworks increasingly center memory; for example, Emotional RAG conditions retrieval on affect to stabilize characters \cite{emotionRAG}. Applied systems carry these mechanics into intimate domains; ConflictLens \cite{chun2025conflictlensllmbasedconflictresolution} for conflict resolution and LoveSims \cite{lovesims} for romantic compatibility, showing agents can scaffold reflection and mirror human communication patterns. Together, these strands shift the view of agents from “chatbots” to interactive, socially intelligent actors and lay the groundwork for dyad-focused simulations such as RELATE-Sim.

\subsection{Social-Science Foundations for Long-Term Matching}
Most dating technologies optimize initial attraction, matching on traits and stated preferences rather than on how partners actually behave together over time \cite{doi:10.1177/1529100612436522, Sharabi2022FindingLO}. Decades of evidence show that what predicts relationship endurance differs from what predicts getting started: long-term stability is better explained by interactional processes—communication quality, responsiveness under stress, conflict repair, and perceived partner commitment—than by demographics or personality profiles \cite{arriaga2006relationship, gottman2014predicts, feelingloved, emotionregulation}. Turning-point theory further argues that couples’ responses to consequential events (e.g., separation, external competition, sacrifice) recalibrate commitment and norms, making these moments especially diagnostic of trajectories \cite{tuningpointinromanticrelationship, perceptionsofturningpoints}. Existing couple interventions can help, but they are typically reactive and their effects can attenuate \cite{roddy}. Together, these strands motivate assessments that move beyond compatibility scores to capture behavior at consequential moments—and ideally give partners a safe way to rehearse responses before high-stakes events arise \cite{doi:10.1177/1529100612436522, whatpredictsfirstdate, tuningpointinromanticrelationship, perceptionsofturningpoints}.

\paragraph{How RELATE-Sim builds on this literature} We operationalize these insights by translating turning-point theory into a category schema (Initial Formation; Relationship Development; Challenges/Tests; Conflict \& Repair; Deepening/Milestones; Other Modern Turning Points) grounded in Baxter \& Bullis \cite{tuningpointinromanticrelationship} and Baxter \& Erbert \cite{perceptionsofturningpoints}, then eliciting and measuring the behaviors known to matter—repair bids, accountability, meta-communication, perspective-taking—rather than relying on static traits \cite{arriaga2006relationship, gottman2014predicts, feelingloved, emotionregulation}. Unlike reactive programs \cite{roddy}, RELATE-Sim provides anticipatory rehearsal and returns actionable signals, aligning HCI design with what social science identifies as the core drivers of long-term endurance.

\section{Turning-Point Taxonomy \& Scenario Design}
\subsection{User Needs and Design Goals}
Our formative interviews and the social-science literature suggest that couples want more than a ``good match.'' They need: 
\begin{itemize}
    \item \textbf{Foresight beyond fit (N1)}: insight into how we handle hard moments, not just whether our interests align \cite{doi:10.1177/1529100612436522, Sharabi2022FindingLO}.
    \item \textbf{Safe rehearsal (N2)}: a low-risk space to practice responses to consequential events and reflect on patterns.
    \item \textbf{Diagnostic clarity (N3)}: concrete, interpretable signals instead of opaque scores.
    \item \textbf{Personalization with comparability (N4)}: scenarios that feel tailored yet still allow apples-to-apples interpretation.
    \item \textbf{Inclusive coverage (N5)}: recognition of diverse cultures, life stages, and modern pressures.
\end{itemize}
\methodname{}'s scene-based simulation system is designed as tools to meet these needs and provides couples an opportunity to reflect, practice, and recognize enduring patterns before they encounter high-stakes situations in real life.

\subsection{Definition and Rationale}
We ground our scenarios in Turning Point Theory \cite{tuningpointinromanticrelationship}, which defines a turning point as “any event or occurrence that is associated with change in a relationship”. Turning points recalibrate norms and commitment—bringing partners closer or pushing them apart—and thus are informative for predicting long-term trajectories \cite{perceptionsofturningpoints}. \methodname{} operationalizes this theory by simulating such events, analyzing how each partner responds, and estimating the commitment of couples.

\subsection{Category Schema}
\label{sec:3:3}
To satisfy \textbf{N1} (foresight) and \textbf{N5} (inclusivity), we organize turning points into six super-categories, each linked to mechanisms emphasized in prior work \cite{tuningpointinromanticrelationship, perceptionsofturningpoints}: \filledcircle{1} \textbf{Initial Formation}: first disclosures, early exclusivity talk, clarifying intentions. \filledcircle{2} \textbf{Relationship Development}: negotiating routines and roles, aligning values, setting boundaries. \filledcircle{3} \textbf{Challenges or Tests}: relocation, financial strain, time scarcity, third-party competition. \filledcircle{4} \textbf{Conflict \& Repair}: grievance episodes, apologies and forgiveness, boundary renegotiation. \filledcircle{5} \textbf{Deepening or Milestones}: cohabitation, engagement/marriage, parenting/caregiving, health events. \filledcircle{6} \textbf{Other Modern Turning Points}: social-media stressors, parasocial/online ties, immigration/visa issues, public visibility of the relationship. This schema is the coverage tool: it ensures we repeatedly test behaviors at the sorts of events that theory says recalibrate commitment.

\subsection{Scenario Bank and Selection}
To meet \textbf{N2} (safe rehearsal) and \textbf{N4} (personalization with comparability), RELATE-Sim maintains an LLM-generated bank of 1,443 generic scenarios, each mapped to a turning-point category from \S\ref{sec:3:3}. Personas used for selection are distilled from de-identified baseline instruments only; no personally identifying details or follow-up outcomes are used. 

For a given dyad, the manager identifies a plausible category using intake signals, personas, simulation history, and current commitment state; it then samples 30 candidate turning points from that category and uses a targeted prompt to select a single best-fit scene idea. The chosen scene is expanded with synthetic yet concrete details (stakes, setting, relevant history, potential third-party involvement) to maximize ecological validity while preserving cross-dyad comparability. 

\subsection{Post-Generation State Inference}
To deliver \textbf{N3} (clear understanding), each scene is labeled with eight interpretable relational states (Appendix \ref{appendix:state_enums}) inferred from the synopsis and planned interaction to track the relationship state: \filledcircle{A} \emph{conflict} - episodes marked by criticism, defensiveness, contempt, or stonewalling that escalate disagreement \filledcircle{B} \emph{repair\_outcome} - moments where partners apologize, forgive, or establish new rituals to resolve tensions \filledcircle{C} \emph{clarity} - Instances when labels, exclusivity, or shared future plans are explicitly negotiated \filledcircle{D} \emph{constraints} - tangible commitments (leases, pets, finances, routines) that raise the cost of exit \filledcircle{E} \emph{alternatives} - signals of interest in rivals, secrecy, or jealousy that highlight external options \filledcircle{F} \emph{transition} - life changes (moves, jobs, distance, schedules) that reshape the relationship context \filledcircle{G} \emph{network} - social approval or disapproval from friends, family, or peers that influences the dyad \filledcircle{H} \emph{breakup\_marker} - clear statements or trial separations indicating relationship dissolution.

An external model (i.e., Scene Master) evaluates evidence for each state and (re)confirms the active turning-point category. These states serve two roles: they steer which turning point comes next (e.g., unresolved conflict triggers a repair-focused episode), and they become plain-language markers users can understand (“clarity moved tacit → explicit,” “two repair attempts; one acknowledged”), while also yielding structured features tied to endurance in prior work \cite{arriaga2006relationship, gottman2014predicts, feelingloved, emotionregulation, tuningpointinromanticrelationship, perceptionsofturningpoints}.

\section{\methodname{}}
\subsection{Agent Design}
\label{sec:agent-design}
\paragraph{Relationship Agents} Agents in \methodname{} represent each partner as a \emph{persona-aligned, memory-grounded, appraisal-driven decision maker} that (i) acts consistently with an extracted persona, (ii) reasons over scene history and long-term memories, (iii) anticipates relational impact at pivotal turning points, and (iv) emits structured actions suitable for downstream analysis.

\paragraph{Agent state and inputs.}
Each agent $i\in{\textsc{A},\textsc{B}}$ maintains a state:
\begin{equation*}
    s_t^i=\langle p^i, \mathcal{M}^i, \mathcal{H}_{1:t}, a^i_t, r^i_t\rangle
\end{equation*}
where $\mathbf{p}^i$ is a 200–300 word persona synopsis derived from Basic Information and interview-like long-form text; $\mathcal{M}^i$ is a long-term memory store of atomic entries with semantic and affect embeddings; $\mathcal{H}_{1:t}$ is the scene history; $\mathbf{a}^i_t\in\mathbb{R}^K$ is an affect appraisal vector (e.g., \emph{joy, sadness, fear, surprise, etc}); and $\mathbf{r}^i_t$ are relationship metrics tracked for analysis (e.g., momentary \emph{dedication}, \emph{alternatives}, \emph{investments} proxies). 
\paragraph{Memory Storage}
Each relationship agent uses a three-layer memory module. Identity Memory (long-term, preloaded) encodes stable traits—personality, background, prior relationship history—and is retrieved semantically to preserve a consistent self-concept. Simulation Memory (episodic, dynamic) accumulates salient experiences across scenes—emotionally charged moments and recurring patterns—also accessed semantically to support continuity and adaptation over time. Scene Memory (short-term, inline) injects the immediate context—what just happened, what was said, and felt—directly into the prompt to keep within-scene responses coherent. Together, these layers enable context-sensitive behavior from moment-to-moment while maintaining a believable long-term arc.

\paragraph{Appraisal and memory retrieval.}
Given scene context $c_t$, the agent computes $\mathbf{a}^i_t=f_{\text{appraise}}(\mathbf{p}^i,\mathcal{H}_{1:t},c_t)$ using an LLM rubric that maps context cues to affect dimensions (Appendix \ref{appendix:emotion_embedding}). 
To capture both semantic and affective proximity between memories and current context, we define a \textit{hybrid-similarity score} that integrates text-based and emotion-based embeddings. Specifically, for a memory $m$, we compute its semantic vector embeddings using the \textit{OpenAI text-embedding-3-small} model, denoted $\textbf{e}_{\text{sem}}(m)$, and its affective embedding using the emotion-embedding representation, denoted $\textbf{e}_{\text{aff}}(m)$. The hybrid score combines these two components through a weighted function, allowing retrieval to privilege memories that are not only topically aligned but also affectively congruent with the ongoing interaction. This design aims to improve recall fidelity in emotionally charged scenes, ensuring that relevant but subtle affective cues are not overshadowed by purely semantic matches. Retrieval selects $k$ salient memories by a \emph{hybrid-similarity} score for the current context $c_t$:
\begin{equation*}
    \text{similarity\_score}(m,c_t)= \cos(\textbf{e}_{\text{sem}}(m), \textbf{e}_{\text{sem}}(c_t))+\lambda \cdot\cos(\textbf{e}_{\text{aff}}(m), \textbf{a}^i_t)
\end{equation*}
returning $\mathcal{R}^i_t=\text{TopK}(\mathcal{M}^i)$. $\lambda$ calibrates affect sensitivity.

Given the option set $\mathcal{O}_t={o_1,\dots,o_M}$ ($M{=}3$–$4$) produced by the Scene Master, the acting agent performs a \emph{single-call} rubric-based selection that is auditable and reproducible. The decision prompt (Appendix \ref{appendix:decision_prompt}) is assembled by fusing five fields: \texttt{agent\_persona}, the agent's persona synopsis \texttt{scene\_context}, a comprehensive log of the current interaction, stakes, and constraints; \texttt{retrieved\_memory}, the top-$k$ episodic snippets; \texttt{internal\_thoughts}, an internal assessment of the context generated during emotional appraisal, \texttt{action\_options}, a compact menu of persona-consistent candidate moves tied to the scene goal. These elements are ordered from situation $\rightarrow$ evidence $\rightarrow$ affect $\rightarrow$ alternatives and inserted into a constrained schema that specifies the required outputs (action, reason, confidence, emotion tags). This structure ensures that generation is context-grounded, memory-aware, and affect-sensitive, while preserving traceability from the emitted action back to salient cues.

\subsection{Environment Design}
\label{sec:scene-manager}
The simulation is coordinated end-to-end by a centralized \emph{Scene Master} (Appendix \ref{appendix:scene_logic}) that acts as a game master: it supervises personas, orchestrates narrative flow, and applies evaluative rubrics. Each scenario is modeled as a discrete \emph{Turning Point} to capture meaningful shifts in the relationship trajectory. To keep scene-to-scene transitions theoretically grounded and interpretable, the system operates over a fixed taxonomy of categories as mentioned in \S\ref{sec:3:3}: \emph{Initial Formation}, \emph{Relationship Development}, \emph{Challenges or Tests}, \emph{Conflict \& Repair}, \emph{Deepening or Milestones}, and \emph{Other Modern Turning Points}.

At the start of each scene, the manager initializes a structured state that serves as the single source of truth for downstream components. This state anchors coherence with prior events and constrains generation to the most relevant goals and tensions

Conditioned on both partners’ personas, the declared conflict, and accumulated scene history, the manager advances the narrative until a natural stopping point—i.e., a juncture at which one partner must act. It then selects the acting partner and constructs a compact option set $\mathcal{O}_t$ of $3$–$4$ \emph{mutually exclusive, single-actor, observable} behaviors that map to distinct relational consequences. Options are grounded in the active turning-point template and the dyadic state $\mathbf{x}_t$, then injected as immutable inputs to the acting agent.

The next turning point is selected via a lightweight heuristic that maps the current \texttt{relationship\_state} to a small set of plausible categories. The manager then conditions a language model on prior summaries, the updated \texttt{relationship\_state}, and a curated list of human-screened candidate turning points; from these, the model selects the most contextually consistent option. The subsequent scene is initialized accordingly, and the cycle repeats.

Architecturally, the Scene Master separates global supervision from local agency. Personas and decision agents operate locally from their goals, while the manager validates options, maintains continuity, and governs transitions and evaluation. State persistence—comprising the scene state (Appendix \ref{appendix:scene_state}), the narrative ledger, rolling summaries, and the \texttt{relationship\_state}—prevents contradictions and repetitive beats across long horizons. Observability is built in through structured logs, decision traces, and versioned checkpoints, enabling auditing and ablation. The design remains extensible: heuristics can be swapped for probabilistic or learned policies, turning-point libraries can be updated without refactoring agents, and governance hooks (e.g., fairness or safety filters) can be applied centrally before enactment.

\section{Experiments}
\label{sec:evaluation}
We test whether final commitment derived from personas and simulated interaction dynamics predicts real-world relationship change over two years.

\subsection{Dataset and Outcome Definition}
\label{sec:data-outcome}
We evaluate on a longitudinal dataset\footnote{The details of the dataset are removed due to anonymity requirements and will be released upon publication.} of $N{=}71$ couples with baseline surveys and a two-year follow-up survey on updated relationship status. Relationship status is recorded at baseline and follow-up (\emph{married, engaged, dating, divorced/broken up}). After cleaning, we define a binary outcome: \emph{improved relationship} (dating/engaged $\rightarrow$ married) versus \emph{stagnant relationship} (status unchanged or worse, including breakup/divorce). This framing enables a clear, externally verifiable target for prediction while preserving ordinal meaning.

\subsection{Persona Synthesis}
\label{sec:persona-synthesis}
We implement a two-stage persona synthesis with GPT-OSS-120b \cite{openai2025gptoss120bgptoss20bmodel}, a large instruction-following language model chosen for its capacity to integrate heterogeneous textual inputs under tight format and style constraints. For each individual, we aggregate multiple instruments: \texttt{ctss} (partner and self description of the subject in conflict), \texttt{ersi} (attention/task/stress traits), \texttt{rpd} (partner’s daily-life view), \texttt{self} (demographic/relationship identity), \texttt{sfn} (partner’s view of friends/interests), and \texttt{vplst} (sources of tension/control). Each instrument is condensed into a short, evidence-linked synopsis using a constrained prompt that privileges concrete, dyad-relevant behaviors and avoids speculation. We then fuse the seven synopses into: (i) a second-person, 200--300 word persona and (ii) a 5--7 rule \emph{playbook} of actionable \textit{if}$\rightarrow$\textit{then} behaviors. This two-stage synthesis standardizes text-based representations for both partners while preserving actor--partner divergences (e.g., \emph{``you report\ldots; your partner experiences\ldots''}). Only baseline data are used.

GPT-OSS-120b is especially effective here because the task demands both breadth (integrating multiple instruments with different lenses) and disciplined controllability (adhering to format, audience, and abstention rules). The model’s long-context reasoning supports cross-document reconciliation—surfacing stable patterns, flagging contradictions, and synthesizing them into actionable second-person language—while constrained prompting and low-variance decoding yield consistent, schema-compliant outputs. Compared to smaller models or manual coding, this approach reduces coder subjectivity, increases reproducibility, and produces richer, behaviorally grounded profiles that translate directly into agent policies (the playbook) and downstream features (e.g., repair tendencies, alternative salience).

Methodologically, the combination of evidence-first summarization and controlled fusion makes the pipeline both powerful and safe. The prompts enforce “evidence or silence,” enabling the model to return neutral/unknown where the record is thin, and the second-person framing operationalizes constructs (attachment, commitment mechanics, communication patterns, regulation style, interdependence) as concrete if→then contingencies. The result is a pair of standardized, high-signal artifacts per individual that are immediately usable for simulation initialization, state updating, and prediction—without requiring any access to follow-up outcomes.

\subsection{Simulation Inference}
Given the computational profile of our simulator—each scene makes 34 distinct LLM calls, most of which contain large context windows—we selected Qwen3-32B \cite{yang2025qwen3technicalreport} as the backbone model for its favorable quality–latency trade-off. In preliminary experiments, 70B-class models(Llama-3.1-70B-instruct \cite{grattafiori2024llama3herdmodels}) provided only modest gains in role fidelity at substantially higher inference latency and cost, whereas smaller models(Llama-3.1-8B-instruct \cite{grattafiori2024llama3herdmodels}) underperformed in long-context recall and persona consistency. Qwen3-32B maintained strong instruction following and narrative coherence while enabling efficient execution on the Lambda Inference API with 16 concurrent simulations. This configuration reduces wall-clock time and stabilizes throughput without sacrificing the reliability required for scene-level reasoning and commitment inference.

\subsection{Baseline Commitment Inference}
Additionally, a baseline commitment is inferred directly from the persona representations rather than from simulated interactions. This baseline measure captures the couple's traits and behavioral patterns at the outset but does not incorporate the dynamic unfolding of scene-level interactions, making it an important point of comparison for evaluating the added value of the simulation.

\section{Results}
\subsection{Simulated Final Commitment}
\label{subsec:evaluation-separation}

We evaluated $n{=}71$ families split into two cohorts and compared (i) a \emph{baseline} commitment inference that uses only partner personas and (ii) a \emph{scene-level simulation} that averages the final commitment across five simulated runs per family. Group means (0-5 scale) are summarized in Table~\ref{tab:means-separation}.

\begin{table}[h]
\centering
\begin{tabular}{lrrrr}
\toprule
\textbf{Cohort} & \textbf{Baseline} & \textbf{RELATE-Sim} & $\boldsymbol{\Delta}$ & $\boldsymbol{\%\Delta}$ \\
\midrule
Decreased-Status & 3.0476 & 2.6060 & $-0.4416$ & $-14.5\%$ \\
Increased-Status & 3.1034 & 2.7759 & $-0.3275$ & $-10.6\%$ \\
\bottomrule
\end{tabular}
\caption{Baseline vs.\ simulation means by cohort. The metric is the predicted commitment score, ranging from 1 to 5. For the decreased-status cohort, the predicted commitment is lower the better. For the increased-status cohort, the predicted commitment is higher the better.\label{tab:means-separation}}
\end{table}

\paragraph{Stronger group discrimination.}
Under the same commitment definition scale, the between-group gap grows from $0.0558$ at baseline ($3.1034-3.0476$) to $0.1699$ in simulation ($2.7759-2.6060$), i.e., a $3.04\times$ larger gap---an increase of $\approx {+}204\%$ in contrast. Thus, the simulator spreads the cohorts farther apart on the 0--5 scale, yielding clearer rank-ordering than the persona-only baseline, whose scores are visibly compressed around ${\sim}3.0$.

\paragraph{Directional recalibration consistent with scenario dynamics.}
Both groups' means shift downward in simulation, but the \textit{decreased-status group} moves more ($-0.4416$, $-14.5\%$) than the \textit{increased-status group} ($-0.3275$, $-10.6\%$). This asymmetric adjustment is expected from injecting scene dynamics: as agents encounter friction points, constraints, and tests of ``we-ness,'' the model penalizes commitment more strongly where simulated interactions expose fragility. The widened gap is driven by greater down-weighting in the lower-commitment cohort, not a uniform rescaling that would preserve baseline compression.

\paragraph{Why this outperforms the baseline.}
\begin{itemize}
\item \textbf{Better separation for prediction.} The expanded gap ($0.170$ vs.\ $0.056$) provides a stronger signal for downstream classifiers or thresholds mapping commitment to outcome categories; minor noise could swamp a $0.056$-point difference, whereas the simulation's tripled separation is more robust.
\item \textbf{Content-aware calibration.} The simulator re-calibrates based on interaction patterns (e.g., repair attempts, protection from alternatives, sliding vs.\ deciding), not a global rescale. Targeted down-adjustment where scenes surface risk factors increases practical informativeness.
\item \textbf{Reduced ceiling/compression effects.} Persona-only estimates cluster near ${\sim}3.0$, masking differences. The simulation broadens the effective range (especially downward), improving sensitivity to relationship-relevant stressors.
\end{itemize}

\subsection{Long-term Label Prediction}
\label{subsec:longterm-labels}

We evaluated a downstream relationship status prediction task using four labels (broken up/divorced, dating, engaged, married) mapped to two end states after two years: \emph{dissolved} vs.\ \emph{sustained}. We compared (i) a \emph{baseline} prompt that observes only the partners’ personas and (ii) a \emph{simulation-aware} prompt that ingests the ordered sequence of scene summaries, plus commitment evaluations and rationales after each scene. In the \emph{simulation-aware} condition, for each family we executed five independent scene simulations, applied the predictor to each run, and selected the modal (most frequent) label across the five outputs as the final prediction. The label prediction task was evaluated on $N=101$ couples (including couples whose relationship status remained constant), and the following results were observed:

\begin{table}[h]
\centering
\caption{Accuracy on two-way end state (dissolved vs.\ sustained).}
\label{tab:longterm-accuracy}
\begin{tabular}{lcccc}
\toprule
\textbf{Model} & \textbf{Correct / N} & \textbf{Accuracy}\\
\midrule
Baseline (personas-only) & $49/101$ & $48.5\%$\\
Simulation-aware & $65/101$ & $64.4\%$ \\
\bottomrule
\end{tabular}
\end{table}

The baseline is statistically indistinguishable from chance (exact binomial vs.\ $0.5$, $p{=}0.84$), whereas the simulation-aware model is reliably above chance (exact binomial vs.\ $0.5$, $p{=}0.005$). The absolute gain is $+15.8$ percentage points ($64.4{-}48.5$), with a normal-approximate 95\% CI of $[+2.3,,+29.3]$ pp. The relative improvement over baseline accuracy is: $\frac{0.6436}{0.4851}-1\approx32.7\%$. Incorporating scene dynamics and the model’s own step-wise commitment plus reasoning yields \emph{content-aware calibration} rather than persona-level heuristics. As conflict patterns, repair attempts, and constraint cues accumulate across scenes, the predictor differentiates trajectories that sustain vs.\ dissolve, overcoming the personas-only baseline’s near-coin-flip ambiguity.
\section{Discussion and Future Work}
\subsection{Current Limitations}
\paragraph{Context window length.}
Our per-turn prompts are intentionally rich: each LLM call may include persona synopses for both partners, scene history and rolling summaries, retrieved autobiographical memories, the current scene state, and appraisal outputs. While necessary for grounded decisions, this long-context setting introduces well-known reliability and efficiency issues.
\begin{itemize}
\item \textbf{Attention diffusion.} As context grows, the model must distribute attention over more tokens, diluting signal. Critical cues compete with background detail, increasing the chance that the model underweights decisive evidence.
\item \textbf{Recency bias.} Transformer attention tends to favor later tokens; earlier but highly relevant details exert less influence as the window lengthens. This produces strong recall for the very recent past and weaker recall for distant-yet-critical facts.
\item \textbf{Error accumulation.} Long contexts often include iterative summaries or intermediate reasoning. Small early misreadings can persist and compound because subsequent turns condition on prior model outputs.
\item \textbf{Computation \& noise.} More tokens increase inference cost and latency. Redundant or noisy detail can blur option ranking and reduce consistency, especially when multiple alternatives are closely matched.
\end{itemize}

\paragraph{Persona summarization and data constraints.}
The source dataset was originally collected to study substance use, not to comprehensively capture romantic personas. As a result, dyad-relevant information is limited and uneven across instruments. Our evaluation does not incorporate long-term autobiographical memory prior to the start of the simulation because no such documents were available. Moreover, condensing seven instruments into brief synopses can drop counter-evidence and minority behaviors; summaries may overweight salient or recent cues and thereby bias downstream selection and appraisal. These constraints limit the granularity and balance of the evidence that feeds persona rules and may attenuate effects tied to less frequent but consequential behaviors.

\paragraph{Group heterogeneity across starting stages.}
Couples enter the study at different stages (dating, cohabiting, married), which challenges direct comparability because “commitment” does not live on a single metric scale across stages. While our outcome definition (dating/engaged~$\rightarrow$~married vs.\ unchanged/worse) offers a clear external target, stage heterogeneity introduces confounds in both baselines and simulations. Interpretations of improvements should therefore be conditioned on starting status and stage-specific opportunities/constraints; residual differences may persist even under stratification or status-adjusted baselines.

\paragraph{Limited temporal dynamics and omitted contexts.}
The simulation models interactions between partners but does not endogenize broader life-course dynamics. Real-world trajectories are shaped by multilevel influences that our current environment does not explicitly represent, including (but not limited to):
\begin{itemize}
\item family-system factors and third-party influence,
\item stress spillover from work/finances/health,
\item traumatic experiences and recovery processes,
\item physical health shocks,
\item macro- and household-level economic strain.
\end{itemize}
Even when such forces are mentioned in personas, the environment does not simulate their timing, magnitude, or persistence; consequently, some mechanisms that drive consolidation or dissolution of commitment may be underrepresented. Together, these limitations suggest that reported effects should be read as evidence for dynamics \emph{within} dyadic interaction frames, not as a full account of the broader ecological realities of long-term relationships.

\subsection{Future Work}
\label{sec:better-design}
\paragraph{Simulation-ready data collection.}
We propose a prospective study that collects \emph{simulation-specific} data with human participants recruited to a common baseline: all dyads begin in the \textbf{dating} stage to standardize early turning points and comparability. For each partner, we gather (i) \emph{self-/partner-reports} tailored to relationship persona modeling (attachment behaviors, maintenance routines, conflict roles, responsiveness, goals), (ii) \emph{self-reported past memories} elicited via short life-history prompts (salient conflicts, repairs, sacrifices, transitions), and (iii) \emph{semi-structured interviews} that surface triggers, soothing moves, and boundary norms. Instruments are designed to yield \emph{evidence-linked} snippets (quote + tag) rather than global trait labels, facilitating persona synthesis and grounding retrieval. All items are timestamped and attributed to actor/partner to preserve divergences and avoid hindsight bias.

\paragraph{Human-in-the-loop evaluation.}
We evaluate ecological validity with \emph{participant-facing simulations}. For a subset of scenes, the Scene Master presents the same $3$--$4$ options to the participant whose persona is instantiated; participants (and optionally their partners) choose the action they would take and provide a one-sentence rationale. We compare (a) \textbf{choice alignment} (agent pick vs.\ human pick), (b) \textbf{ranking agreement} (agent vs.\ human option ordering), and (c) \textbf{rationale overlap} (shared cues/memories cited). Beyond accuracy, we solicit \emph{feedback at decision points} (missing options, misread persona cues, unsafe framings) and incorporate these annotations to refine option curation and rubric weights. This loop yields both quantitative agreement metrics and qualitative error analyses that directly inform prompt and policy updates.

\paragraph{External stressors and spillover.}
To move beyond purely dyadic dynamics, we extend the environment with \emph{exogenous shock modules} (e.g., family interference, job loss, health scare, unexpected caregiving, housing/financial strain). Each module is encoded as a structured event with: trigger conditions, duration, intensity, and \emph{spillover functions} that perturb state variables (tension, trust, clarity, constraints) over time. Shocks may enter stochastically (hazard-based) or via scripted injections aligned with participant reports. The Scene Master surfaces these events as turning-point frames, enabling agents to display resilience, planning, and support-seeking behaviors that better approximate real-world trajectories.

\paragraph{Scalability and personalization platform.}
We envision a multi-tenant platform for \emph{personalized relationship simulation} oriented toward long-term fit rather than first-date matching. Users contribute their personas (surveys, brief interviews, optional memory snippets) and can run \emph{one-to-many} simulations: their agent interacts with a pool of potential partners under common scenario sets and shared stressor libraries. Simulations run concurrently, logging behavioral features (repair rates, clarity gains, escalation avoidance), outcome proxies (commitment trajectories), and robustness under shocks. A \emph{compatibility score} is then derived from cross-simulation performance (e.g., probability of stable deepening across milestones, calibration of expectations, avoidance of high-risk cycles), subject to fairness/safety screening and transparent explanations. This architecture supports iterative onboarding (start simple, add memories later), progressive disclosure (users see how profiles influence choices), and research-grade telemetry (versioned prompts, seeds, and decision traces).

\subsection{Ethical Statement}
We used a longitudinal dataset collected under a U.S. federal grant; details are withheld for anonymous review and will be revealed upon publication. All study protocols were approved by the authors’ Institutional Review Boards. Moreover, AI has been used to assist paper writing, such as polishing languages for better expressivity.

\begin{acks}
We gratefully acknowledge the contribution of the Project Alliance staff, Portland Public Schools, and the participating families. The research reported in this paper was supported by grants from the National Institute of Drug Abuse (DA07031), and the National Institute on Alcoholism and Alcohol Abuse (AA022071), both awarded to T. Ha. The content is solely the responsibility of the authors and does not necessarily reflect the official views of the National Institute on Drug Abuse and the National Institute on Alcoholism and Alcohol Abuse.
\end{acks}

\newpage
\bibliographystyle{ACM-Reference-Format}
\bibliography{bib}


\begin{thebibliography}{23}


\ifx \showCODEN    \undefined \def \showCODEN     #1{\unskip}     \fi
\ifx \showISBNx    \undefined \def \showISBNx     #1{\unskip}     \fi
\ifx \showISBNxiii \undefined \def \showISBNxiii  #1{\unskip}     \fi
\ifx \showISSN     \undefined \def \showISSN      #1{\unskip}     \fi
\ifx \showLCCN     \undefined \def \showLCCN      #1{\unskip}     \fi
\ifx \shownote     \undefined \def \shownote      #1{#1}          \fi
\ifx \showarticletitle \undefined \def \showarticletitle #1{#1}   \fi
\ifx \showURL      \undefined \def \showURL       {\relax}        \fi
\providecommand\bibfield[2]{#2}
\providecommand\bibinfo[2]{#2}
\providecommand\natexlab[1]{#1}
\providecommand\showeprint[2][]{arXiv:#2}

\bibitem[Arriaga et~al\mbox{.}(2006)]%
        {arriaga2006relationship}
\bibfield{author}{\bibinfo{person}{Ximena~B Arriaga}, \bibinfo{person}{Jason~T Reed}, \bibinfo{person}{Wind Goodfriend}, {and} \bibinfo{person}{Christopher~R Agnew}.} \bibinfo{year}{2006}\natexlab{}.
\newblock \showarticletitle{Relationship perceptions and persistence: Do fluctuations in perceived partner commitment undermine dating relationships?}
\newblock \bibinfo{journal}{\emph{Journal of personality and social psychology}} \bibinfo{volume}{91}, \bibinfo{number}{6} (\bibinfo{year}{2006}), \bibinfo{pages}{1045}.
\newblock
\urldef\tempurl%
\url{https://psycnet.apa.org/buy/2006-21634-005}
\showURL{%
\tempurl}


\bibitem[Ashery et~al\mbox{.}(2025)]%
        {city35211}
\bibfield{author}{\bibinfo{person}{A.~F. Ashery}, \bibinfo{person}{L.~M. Aiello}, {and} \bibinfo{person}{A. Baronchelli}.} \bibinfo{year}{2025}\natexlab{}.
\newblock \showarticletitle{Emergent social conventions and collective bias in LLM populations}.
\newblock \bibinfo{journal}{\emph{Science Advances}} \bibinfo{volume}{11}, \bibinfo{number}{20} (\bibinfo{date}{May} \bibinfo{year}{2025}), \bibinfo{pages}{eadu9368--}.
\newblock
\showISSN{2375-2548}
\href{https://doi.org/10.1126/sciadv.adu9368}{doi:\nolinkurl{10.1126/sciadv.adu9368}}
\newblock
\shownote{Copyright {\copyright} 2025 the Authors, some rights reserved; exclusive licensee American Association for the Advancement of Science. no claim to original U.S. Government Works. distributed under a creative commons Attribution noncommercial license 4.0 (cc BY- nc).}.


\bibitem[Baxter and Bullis(1986)]%
        {tuningpointinromanticrelationship}
\bibfield{author}{\bibinfo{person}{Leslie~A. Baxter} {and} \bibinfo{person}{Connie Bullis}.} \bibinfo{year}{1986}\natexlab{}.
\newblock \showarticletitle{Turning Points in Developing Romantic Relationships}.
\newblock \bibinfo{journal}{\emph{Human Communication Research}} \bibinfo{volume}{12}, \bibinfo{number}{4} (\bibinfo{date}{03} \bibinfo{year}{1986}), \bibinfo{pages}{469--493}.
\newblock
\showISSN{0360-3989}
\showeprint{https://academic.oup.com/hcr/article-pdf/12/4/469/22342512/jhumcom0469.pdf}
\href{https://doi.org/10.1111/j.1468-2958.1986.tb00088.x}{doi:\nolinkurl{10.1111/j.1468-2958.1986.tb00088.x}}


\bibitem[Baxter and Erbert(1999)]%
        {perceptionsofturningpoints}
\bibfield{author}{\bibinfo{person}{Leslie~A. Baxter} {and} \bibinfo{person}{Larry~A. Erbert}.} \bibinfo{year}{1999}\natexlab{}.
\newblock \showarticletitle{Perceptions of Dialectical Contradictions in Turning Points of Development in Heterosexual Romantic Relationships}.
\newblock \bibinfo{journal}{\emph{Journal of Social and Personal Relationships}} \bibinfo{volume}{16}, \bibinfo{number}{5} (\bibinfo{year}{1999}), \bibinfo{pages}{547--569}.
\newblock
\showeprint{https://doi.org/10.1177/0265407599165001}
\href{https://doi.org/10.1177/0265407599165001}{doi:\nolinkurl{10.1177/0265407599165001}}


\bibitem[Chun et~al\mbox{.}(2025)]%
        {chun2025conflictlensllmbasedconflictresolution}
\bibfield{author}{\bibinfo{person}{Jiwon Chun}, \bibinfo{person}{Gefei Zhang}, {and} \bibinfo{person}{Meng Xia}.} \bibinfo{year}{2025}\natexlab{}.
\newblock \bibinfo{title}{ConflictLens: LLM-Based Conflict Resolution Training in Romantic Relationship}.
\newblock
\showeprint[arxiv]{2505.11715}~[cs.HC]
\urldef\tempurl%
\url{https://arxiv.org/abs/2505.11715}
\showURL{%
\tempurl}


\bibitem[Finkel et~al\mbox{.}(2012)]%
        {doi:10.1177/1529100612436522}
\bibfield{author}{\bibinfo{person}{Eli~J. Finkel}, \bibinfo{person}{Paul~W. Eastwick}, \bibinfo{person}{Benjamin~R. Karney}, \bibinfo{person}{Harry~T. Reis}, {and} \bibinfo{person}{Susan Sprecher}.} \bibinfo{year}{2012}\natexlab{}.
\newblock \showarticletitle{Online Dating: A Critical Analysis From the Perspective of Psychological Science}.
\newblock \bibinfo{journal}{\emph{Psychological Science in the Public Interest}} \bibinfo{volume}{13}, \bibinfo{number}{1} (\bibinfo{year}{2012}), \bibinfo{pages}{3--66}.
\newblock
\showeprint{https://doi.org/10.1177/1529100612436522}
\href{https://doi.org/10.1177/1529100612436522}{doi:\nolinkurl{10.1177/1529100612436522}}
\newblock
\shownote{PMID: 26173279}.


\bibitem[Gottman(2014)]%
        {gottman2014predicts}
\bibfield{author}{\bibinfo{person}{J.M. Gottman}.} \bibinfo{year}{2014}\natexlab{}.
\newblock \bibinfo{booktitle}{\emph{What Predicts Divorce?: The Relationship Between Marital Processes and Marital Outcomes}}.
\newblock \bibinfo{publisher}{Taylor \& Francis}.
\newblock
\showISBNx{9781317781646}
\urldef\tempurl%
\url{https://books.google.com/books?id=ziABAwAAQBAJ}
\showURL{%
\tempurl}


\bibitem[Grattafiori et~al\mbox{.}(2024)]%
        {grattafiori2024llama3herdmodels}
\bibfield{author}{\bibinfo{person}{Aaron Grattafiori}, \bibinfo{person}{Abhimanyu Dubey}, \bibinfo{person}{Abhinav Jauhri}, \bibinfo{person}{Abhinav Pandey}, \bibinfo{person}{Abhishek Kadian}, \bibinfo{person}{Ahmad Al-Dahle}, \bibinfo{person}{Aiesha Letman}, \bibinfo{person}{Akhil Mathur}, \bibinfo{person}{Alan Schelten}, \bibinfo{person}{Alex Vaughan}, \bibinfo{person}{Amy Yang}, \bibinfo{person}{Angela Fan}, \bibinfo{person}{Anirudh Goyal}, \bibinfo{person}{Anthony Hartshorn}, \bibinfo{person}{Aobo Yang}, \bibinfo{person}{Archi Mitra}, \bibinfo{person}{Archie Sravankumar}, \bibinfo{person}{Artem Korenev}, \bibinfo{person}{Arthur Hinsvark}, \bibinfo{person}{Arun Rao}, \bibinfo{person}{Aston Zhang}, \bibinfo{person}{Aurelien Rodriguez}, \bibinfo{person}{Austen Gregerson}, \bibinfo{person}{Ava Spataru}, \bibinfo{person}{Baptiste Roziere}, \bibinfo{person}{Bethany Biron}, \bibinfo{person}{Binh Tang}, \bibinfo{person}{Bobbie Chern}, \bibinfo{person}{Charlotte Caucheteux}, \bibinfo{person}{Chaya Nayak},
  \bibinfo{person}{Chloe Bi}, \bibinfo{person}{Chris Marra}, \bibinfo{person}{Chris McConnell}, \bibinfo{person}{Christian Keller}, \bibinfo{person}{Christophe Touret}, \bibinfo{person}{Chunyang Wu}, \bibinfo{person}{Corinne Wong}, \bibinfo{person}{Cristian~Canton Ferrer}, \bibinfo{person}{Cyrus Nikolaidis}, \bibinfo{person}{Damien Allonsius}, \bibinfo{person}{Daniel Song}, \bibinfo{person}{Danielle Pintz}, \bibinfo{person}{Danny Livshits}, \bibinfo{person}{Danny Wyatt}, \bibinfo{person}{David Esiobu}, \bibinfo{person}{Dhruv Choudhary}, \bibinfo{person}{Dhruv Mahajan}, \bibinfo{person}{Diego Garcia-Olano}, \bibinfo{person}{Diego Perino}, \bibinfo{person}{Dieuwke Hupkes}, \bibinfo{person}{Egor Lakomkin}, \bibinfo{person}{Ehab AlBadawy}, \bibinfo{person}{Elina Lobanova}, \bibinfo{person}{Emily Dinan}, \bibinfo{person}{Eric~Michael Smith}, \bibinfo{person}{Filip Radenovic}, \bibinfo{person}{Francisco Guzmán}, \bibinfo{person}{Frank Zhang}, \bibinfo{person}{Gabriel Synnaeve}, \bibinfo{person}{Gabrielle Lee},
  \bibinfo{person}{Georgia~Lewis Anderson}, \bibinfo{person}{Govind Thattai}, \bibinfo{person}{Graeme Nail}, \bibinfo{person}{Gregoire Mialon}, \bibinfo{person}{Guan Pang}, \bibinfo{person}{Guillem Cucurell}, \bibinfo{person}{Hailey Nguyen}, \bibinfo{person}{Hannah Korevaar}, \bibinfo{person}{Hu Xu}, \bibinfo{person}{Hugo Touvron}, \bibinfo{person}{Iliyan Zarov}, \bibinfo{person}{Imanol~Arrieta Ibarra}, \bibinfo{person}{Isabel Kloumann}, \bibinfo{person}{Ishan Misra}, \bibinfo{person}{Ivan Evtimov}, \bibinfo{person}{Jack Zhang}, \bibinfo{person}{Jade Copet}, \bibinfo{person}{Jaewon Lee}, \bibinfo{person}{Jan Geffert}, \bibinfo{person}{Jana Vranes}, \bibinfo{person}{Jason Park}, \bibinfo{person}{Jay Mahadeokar}, \bibinfo{person}{Jeet Shah}, \bibinfo{person}{Jelmer van~der Linde}, \bibinfo{person}{Jennifer Billock}, \bibinfo{person}{Jenny Hong}, \bibinfo{person}{Jenya Lee}, \bibinfo{person}{Jeremy Fu}, \bibinfo{person}{Jianfeng Chi}, \bibinfo{person}{Jianyu Huang}, \bibinfo{person}{Jiawen Liu},
  \bibinfo{person}{Jie Wang}, \bibinfo{person}{Jiecao Yu}, \bibinfo{person}{Joanna Bitton}, \bibinfo{person}{Joe Spisak}, \bibinfo{person}{Jongsoo Park}, \bibinfo{person}{Joseph Rocca}, \bibinfo{person}{Joshua Johnstun}, \bibinfo{person}{Joshua Saxe}, \bibinfo{person}{Junteng Jia}, \bibinfo{person}{Kalyan~Vasuden Alwala}, \bibinfo{person}{Karthik Prasad}, \bibinfo{person}{Kartikeya Upasani}, \bibinfo{person}{Kate Plawiak}, \bibinfo{person}{Ke Li}, \bibinfo{person}{Kenneth Heafield}, \bibinfo{person}{Kevin Stone}, \bibinfo{person}{Khalid El-Arini}, \bibinfo{person}{Krithika Iyer}, \bibinfo{person}{Kshitiz Malik}, \bibinfo{person}{Kuenley Chiu}, \bibinfo{person}{Kunal Bhalla}, \bibinfo{person}{Kushal Lakhotia}, \bibinfo{person}{Lauren Rantala-Yeary}, \bibinfo{person}{Laurens van~der Maaten}, \bibinfo{person}{Lawrence Chen}, \bibinfo{person}{Liang Tan}, \bibinfo{person}{Liz Jenkins}, \bibinfo{person}{Louis Martin}, \bibinfo{person}{Lovish Madaan}, \bibinfo{person}{Lubo Malo}, \bibinfo{person}{Lukas Blecher},
  \bibinfo{person}{Lukas Landzaat}, \bibinfo{person}{Luke de Oliveira}, \bibinfo{person}{Madeline Muzzi}, \bibinfo{person}{Mahesh Pasupuleti}, \bibinfo{person}{Mannat Singh}, \bibinfo{person}{Manohar Paluri}, \bibinfo{person}{Marcin Kardas}, \bibinfo{person}{Maria Tsimpoukelli}, \bibinfo{person}{Mathew Oldham}, \bibinfo{person}{Mathieu Rita}, \bibinfo{person}{Maya Pavlova}, \bibinfo{person}{Melanie Kambadur}, \bibinfo{person}{Mike Lewis}, \bibinfo{person}{Min Si}, \bibinfo{person}{Mitesh~Kumar Singh}, \bibinfo{person}{Mona Hassan}, \bibinfo{person}{Naman Goyal}, \bibinfo{person}{Narjes Torabi}, \bibinfo{person}{Nikolay Bashlykov}, \bibinfo{person}{Nikolay Bogoychev}, \bibinfo{person}{Niladri Chatterji}, \bibinfo{person}{Ning Zhang}, \bibinfo{person}{Olivier Duchenne}, \bibinfo{person}{Onur Çelebi}, \bibinfo{person}{Patrick Alrassy}, \bibinfo{person}{Pengchuan Zhang}, \bibinfo{person}{Pengwei Li}, \bibinfo{person}{Petar Vasic}, \bibinfo{person}{Peter Weng}, \bibinfo{person}{Prajjwal Bhargava},
  \bibinfo{person}{Pratik Dubal}, \bibinfo{person}{Praveen Krishnan}, \bibinfo{person}{Punit~Singh Koura}, \bibinfo{person}{Puxin Xu}, \bibinfo{person}{Qing He}, \bibinfo{person}{Qingxiao Dong}, \bibinfo{person}{Ragavan Srinivasan}, \bibinfo{person}{Raj Ganapathy}, \bibinfo{person}{Ramon Calderer}, \bibinfo{person}{Ricardo~Silveira Cabral}, \bibinfo{person}{Robert Stojnic}, \bibinfo{person}{Roberta Raileanu}, \bibinfo{person}{Rohan Maheswari}, \bibinfo{person}{Rohit Girdhar}, \bibinfo{person}{Rohit Patel}, \bibinfo{person}{Romain Sauvestre}, \bibinfo{person}{Ronnie Polidoro}, \bibinfo{person}{Roshan Sumbaly}, \bibinfo{person}{Ross Taylor}, \bibinfo{person}{Ruan Silva}, \bibinfo{person}{Rui Hou}, \bibinfo{person}{Rui Wang}, \bibinfo{person}{Saghar Hosseini}, \bibinfo{person}{Sahana Chennabasappa}, \bibinfo{person}{Sanjay Singh}, \bibinfo{person}{Sean Bell}, \bibinfo{person}{Seohyun~Sonia Kim}, \bibinfo{person}{Sergey Edunov}, \bibinfo{person}{Shaoliang Nie}, \bibinfo{person}{Sharan Narang},
  \bibinfo{person}{Sharath Raparthy}, \bibinfo{person}{Sheng Shen}, \bibinfo{person}{Shengye Wan}, \bibinfo{person}{Shruti Bhosale}, \bibinfo{person}{Shun Zhang}, \bibinfo{person}{Simon Vandenhende}, \bibinfo{person}{Soumya Batra}, \bibinfo{person}{Spencer Whitman}, \bibinfo{person}{Sten Sootla}, \bibinfo{person}{Stephane Collot}, \bibinfo{person}{Suchin Gururangan}, \bibinfo{person}{Sydney Borodinsky}, \bibinfo{person}{Tamar Herman}, \bibinfo{person}{Tara Fowler}, \bibinfo{person}{Tarek Sheasha}, \bibinfo{person}{Thomas Georgiou}, \bibinfo{person}{Thomas Scialom}, \bibinfo{person}{Tobias Speckbacher}, \bibinfo{person}{Todor Mihaylov}, \bibinfo{person}{Tong Xiao}, \bibinfo{person}{Ujjwal Karn}, \bibinfo{person}{Vedanuj Goswami}, \bibinfo{person}{Vibhor Gupta}, \bibinfo{person}{Vignesh Ramanathan}, \bibinfo{person}{Viktor Kerkez}, \bibinfo{person}{Vincent Gonguet}, \bibinfo{person}{Virginie Do}, \bibinfo{person}{Vish Vogeti}, \bibinfo{person}{Vítor Albiero}, \bibinfo{person}{Vladan Petrovic},
  \bibinfo{person}{Weiwei Chu}, \bibinfo{person}{Wenhan Xiong}, \bibinfo{person}{Wenyin Fu}, \bibinfo{person}{Whitney Meers}, \bibinfo{person}{Xavier Martinet}, \bibinfo{person}{Xiaodong Wang}, \bibinfo{person}{Xiaofang Wang}, \bibinfo{person}{Xiaoqing~Ellen Tan}, \bibinfo{person}{Xide Xia}, \bibinfo{person}{Xinfeng Xie}, \bibinfo{person}{Xuchao Jia}, \bibinfo{person}{Xuewei Wang}, \bibinfo{person}{Yaelle Goldschlag}, \bibinfo{person}{Yashesh Gaur}, \bibinfo{person}{Yasmine Babaei}, \bibinfo{person}{Yi Wen}, \bibinfo{person}{Yiwen Song}, \bibinfo{person}{Yuchen Zhang}, \bibinfo{person}{Yue Li}, \bibinfo{person}{Yuning Mao}, \bibinfo{person}{Zacharie~Delpierre Coudert}, \bibinfo{person}{Zheng Yan}, \bibinfo{person}{Zhengxing Chen}, \bibinfo{person}{Zoe Papakipos}, \bibinfo{person}{Aaditya Singh}, \bibinfo{person}{Aayushi Srivastava}, \bibinfo{person}{Abha Jain}, \bibinfo{person}{Adam Kelsey}, \bibinfo{person}{Adam Shajnfeld}, \bibinfo{person}{Adithya Gangidi}, \bibinfo{person}{Adolfo Victoria},
  \bibinfo{person}{Ahuva Goldstand}, \bibinfo{person}{Ajay Menon}, \bibinfo{person}{Ajay Sharma}, \bibinfo{person}{Alex Boesenberg}, \bibinfo{person}{Alexei Baevski}, \bibinfo{person}{Allie Feinstein}, \bibinfo{person}{Amanda Kallet}, \bibinfo{person}{Amit Sangani}, \bibinfo{person}{Amos Teo}, \bibinfo{person}{Anam Yunus}, \bibinfo{person}{Andrei Lupu}, \bibinfo{person}{Andres Alvarado}, \bibinfo{person}{Andrew Caples}, \bibinfo{person}{Andrew Gu}, \bibinfo{person}{Andrew Ho}, \bibinfo{person}{Andrew Poulton}, \bibinfo{person}{Andrew Ryan}, \bibinfo{person}{Ankit Ramchandani}, \bibinfo{person}{Annie Dong}, \bibinfo{person}{Annie Franco}, \bibinfo{person}{Anuj Goyal}, \bibinfo{person}{Aparajita Saraf}, \bibinfo{person}{Arkabandhu Chowdhury}, \bibinfo{person}{Ashley Gabriel}, \bibinfo{person}{Ashwin Bharambe}, \bibinfo{person}{Assaf Eisenman}, \bibinfo{person}{Azadeh Yazdan}, \bibinfo{person}{Beau James}, \bibinfo{person}{Ben Maurer}, \bibinfo{person}{Benjamin Leonhardi}, \bibinfo{person}{Bernie Huang},
  \bibinfo{person}{Beth Loyd}, \bibinfo{person}{Beto~De Paola}, \bibinfo{person}{Bhargavi Paranjape}, \bibinfo{person}{Bing Liu}, \bibinfo{person}{Bo Wu}, \bibinfo{person}{Boyu Ni}, \bibinfo{person}{Braden Hancock}, \bibinfo{person}{Bram Wasti}, \bibinfo{person}{Brandon Spence}, \bibinfo{person}{Brani Stojkovic}, \bibinfo{person}{Brian Gamido}, \bibinfo{person}{Britt Montalvo}, \bibinfo{person}{Carl Parker}, \bibinfo{person}{Carly Burton}, \bibinfo{person}{Catalina Mejia}, \bibinfo{person}{Ce Liu}, \bibinfo{person}{Changhan Wang}, \bibinfo{person}{Changkyu Kim}, \bibinfo{person}{Chao Zhou}, \bibinfo{person}{Chester Hu}, \bibinfo{person}{Ching-Hsiang Chu}, \bibinfo{person}{Chris Cai}, \bibinfo{person}{Chris Tindal}, \bibinfo{person}{Christoph Feichtenhofer}, \bibinfo{person}{Cynthia Gao}, \bibinfo{person}{Damon Civin}, \bibinfo{person}{Dana Beaty}, \bibinfo{person}{Daniel Kreymer}, \bibinfo{person}{Daniel Li}, \bibinfo{person}{David Adkins}, \bibinfo{person}{David Xu}, \bibinfo{person}{Davide Testuggine},
  \bibinfo{person}{Delia David}, \bibinfo{person}{Devi Parikh}, \bibinfo{person}{Diana Liskovich}, \bibinfo{person}{Didem Foss}, \bibinfo{person}{Dingkang Wang}, \bibinfo{person}{Duc Le}, \bibinfo{person}{Dustin Holland}, \bibinfo{person}{Edward Dowling}, \bibinfo{person}{Eissa Jamil}, \bibinfo{person}{Elaine Montgomery}, \bibinfo{person}{Eleonora Presani}, \bibinfo{person}{Emily Hahn}, \bibinfo{person}{Emily Wood}, \bibinfo{person}{Eric-Tuan Le}, \bibinfo{person}{Erik Brinkman}, \bibinfo{person}{Esteban Arcaute}, \bibinfo{person}{Evan Dunbar}, \bibinfo{person}{Evan Smothers}, \bibinfo{person}{Fei Sun}, \bibinfo{person}{Felix Kreuk}, \bibinfo{person}{Feng Tian}, \bibinfo{person}{Filippos Kokkinos}, \bibinfo{person}{Firat Ozgenel}, \bibinfo{person}{Francesco Caggioni}, \bibinfo{person}{Frank Kanayet}, \bibinfo{person}{Frank Seide}, \bibinfo{person}{Gabriela~Medina Florez}, \bibinfo{person}{Gabriella Schwarz}, \bibinfo{person}{Gada Badeer}, \bibinfo{person}{Georgia Swee}, \bibinfo{person}{Gil Halpern},
  \bibinfo{person}{Grant Herman}, \bibinfo{person}{Grigory Sizov}, \bibinfo{person}{Guangyi}, \bibinfo{person}{Zhang}, \bibinfo{person}{Guna Lakshminarayanan}, \bibinfo{person}{Hakan Inan}, \bibinfo{person}{Hamid Shojanazeri}, \bibinfo{person}{Han Zou}, \bibinfo{person}{Hannah Wang}, \bibinfo{person}{Hanwen Zha}, \bibinfo{person}{Haroun Habeeb}, \bibinfo{person}{Harrison Rudolph}, \bibinfo{person}{Helen Suk}, \bibinfo{person}{Henry Aspegren}, \bibinfo{person}{Hunter Goldman}, \bibinfo{person}{Hongyuan Zhan}, \bibinfo{person}{Ibrahim Damlaj}, \bibinfo{person}{Igor Molybog}, \bibinfo{person}{Igor Tufanov}, \bibinfo{person}{Ilias Leontiadis}, \bibinfo{person}{Irina-Elena Veliche}, \bibinfo{person}{Itai Gat}, \bibinfo{person}{Jake Weissman}, \bibinfo{person}{James Geboski}, \bibinfo{person}{James Kohli}, \bibinfo{person}{Janice Lam}, \bibinfo{person}{Japhet Asher}, \bibinfo{person}{Jean-Baptiste Gaya}, \bibinfo{person}{Jeff Marcus}, \bibinfo{person}{Jeff Tang}, \bibinfo{person}{Jennifer Chan},
  \bibinfo{person}{Jenny Zhen}, \bibinfo{person}{Jeremy Reizenstein}, \bibinfo{person}{Jeremy Teboul}, \bibinfo{person}{Jessica Zhong}, \bibinfo{person}{Jian Jin}, \bibinfo{person}{Jingyi Yang}, \bibinfo{person}{Joe Cummings}, \bibinfo{person}{Jon Carvill}, \bibinfo{person}{Jon Shepard}, \bibinfo{person}{Jonathan McPhie}, \bibinfo{person}{Jonathan Torres}, \bibinfo{person}{Josh Ginsburg}, \bibinfo{person}{Junjie Wang}, \bibinfo{person}{Kai Wu}, \bibinfo{person}{Kam~Hou U}, \bibinfo{person}{Karan Saxena}, \bibinfo{person}{Kartikay Khandelwal}, \bibinfo{person}{Katayoun Zand}, \bibinfo{person}{Kathy Matosich}, \bibinfo{person}{Kaushik Veeraraghavan}, \bibinfo{person}{Kelly Michelena}, \bibinfo{person}{Keqian Li}, \bibinfo{person}{Kiran Jagadeesh}, \bibinfo{person}{Kun Huang}, \bibinfo{person}{Kunal Chawla}, \bibinfo{person}{Kyle Huang}, \bibinfo{person}{Lailin Chen}, \bibinfo{person}{Lakshya Garg}, \bibinfo{person}{Lavender A}, \bibinfo{person}{Leandro Silva}, \bibinfo{person}{Lee Bell}, \bibinfo{person}{Lei
  Zhang}, \bibinfo{person}{Liangpeng Guo}, \bibinfo{person}{Licheng Yu}, \bibinfo{person}{Liron Moshkovich}, \bibinfo{person}{Luca Wehrstedt}, \bibinfo{person}{Madian Khabsa}, \bibinfo{person}{Manav Avalani}, \bibinfo{person}{Manish Bhatt}, \bibinfo{person}{Martynas Mankus}, \bibinfo{person}{Matan Hasson}, \bibinfo{person}{Matthew Lennie}, \bibinfo{person}{Matthias Reso}, \bibinfo{person}{Maxim Groshev}, \bibinfo{person}{Maxim Naumov}, \bibinfo{person}{Maya Lathi}, \bibinfo{person}{Meghan Keneally}, \bibinfo{person}{Miao Liu}, \bibinfo{person}{Michael~L. Seltzer}, \bibinfo{person}{Michal Valko}, \bibinfo{person}{Michelle Restrepo}, \bibinfo{person}{Mihir Patel}, \bibinfo{person}{Mik Vyatskov}, \bibinfo{person}{Mikayel Samvelyan}, \bibinfo{person}{Mike Clark}, \bibinfo{person}{Mike Macey}, \bibinfo{person}{Mike Wang}, \bibinfo{person}{Miquel~Jubert Hermoso}, \bibinfo{person}{Mo Metanat}, \bibinfo{person}{Mohammad Rastegari}, \bibinfo{person}{Munish Bansal}, \bibinfo{person}{Nandhini Santhanam},
  \bibinfo{person}{Natascha Parks}, \bibinfo{person}{Natasha White}, \bibinfo{person}{Navyata Bawa}, \bibinfo{person}{Nayan Singhal}, \bibinfo{person}{Nick Egebo}, \bibinfo{person}{Nicolas Usunier}, \bibinfo{person}{Nikhil Mehta}, \bibinfo{person}{Nikolay~Pavlovich Laptev}, \bibinfo{person}{Ning Dong}, \bibinfo{person}{Norman Cheng}, \bibinfo{person}{Oleg Chernoguz}, \bibinfo{person}{Olivia Hart}, \bibinfo{person}{Omkar Salpekar}, \bibinfo{person}{Ozlem Kalinli}, \bibinfo{person}{Parkin Kent}, \bibinfo{person}{Parth Parekh}, \bibinfo{person}{Paul Saab}, \bibinfo{person}{Pavan Balaji}, \bibinfo{person}{Pedro Rittner}, \bibinfo{person}{Philip Bontrager}, \bibinfo{person}{Pierre Roux}, \bibinfo{person}{Piotr Dollar}, \bibinfo{person}{Polina Zvyagina}, \bibinfo{person}{Prashant Ratanchandani}, \bibinfo{person}{Pritish Yuvraj}, \bibinfo{person}{Qian Liang}, \bibinfo{person}{Rachad Alao}, \bibinfo{person}{Rachel Rodriguez}, \bibinfo{person}{Rafi Ayub}, \bibinfo{person}{Raghotham Murthy}, \bibinfo{person}{Raghu
  Nayani}, \bibinfo{person}{Rahul Mitra}, \bibinfo{person}{Rangaprabhu Parthasarathy}, \bibinfo{person}{Raymond Li}, \bibinfo{person}{Rebekkah Hogan}, \bibinfo{person}{Robin Battey}, \bibinfo{person}{Rocky Wang}, \bibinfo{person}{Russ Howes}, \bibinfo{person}{Ruty Rinott}, \bibinfo{person}{Sachin Mehta}, \bibinfo{person}{Sachin Siby}, \bibinfo{person}{Sai~Jayesh Bondu}, \bibinfo{person}{Samyak Datta}, \bibinfo{person}{Sara Chugh}, \bibinfo{person}{Sara Hunt}, \bibinfo{person}{Sargun Dhillon}, \bibinfo{person}{Sasha Sidorov}, \bibinfo{person}{Satadru Pan}, \bibinfo{person}{Saurabh Mahajan}, \bibinfo{person}{Saurabh Verma}, \bibinfo{person}{Seiji Yamamoto}, \bibinfo{person}{Sharadh Ramaswamy}, \bibinfo{person}{Shaun Lindsay}, \bibinfo{person}{Shaun Lindsay}, \bibinfo{person}{Sheng Feng}, \bibinfo{person}{Shenghao Lin}, \bibinfo{person}{Shengxin~Cindy Zha}, \bibinfo{person}{Shishir Patil}, \bibinfo{person}{Shiva Shankar}, \bibinfo{person}{Shuqiang Zhang}, \bibinfo{person}{Shuqiang Zhang}, \bibinfo{person}{Sinong
  Wang}, \bibinfo{person}{Sneha Agarwal}, \bibinfo{person}{Soji Sajuyigbe}, \bibinfo{person}{Soumith Chintala}, \bibinfo{person}{Stephanie Max}, \bibinfo{person}{Stephen Chen}, \bibinfo{person}{Steve Kehoe}, \bibinfo{person}{Steve Satterfield}, \bibinfo{person}{Sudarshan Govindaprasad}, \bibinfo{person}{Sumit Gupta}, \bibinfo{person}{Summer Deng}, \bibinfo{person}{Sungmin Cho}, \bibinfo{person}{Sunny Virk}, \bibinfo{person}{Suraj Subramanian}, \bibinfo{person}{Sy Choudhury}, \bibinfo{person}{Sydney Goldman}, \bibinfo{person}{Tal Remez}, \bibinfo{person}{Tamar Glaser}, \bibinfo{person}{Tamara Best}, \bibinfo{person}{Thilo Koehler}, \bibinfo{person}{Thomas Robinson}, \bibinfo{person}{Tianhe Li}, \bibinfo{person}{Tianjun Zhang}, \bibinfo{person}{Tim Matthews}, \bibinfo{person}{Timothy Chou}, \bibinfo{person}{Tzook Shaked}, \bibinfo{person}{Varun Vontimitta}, \bibinfo{person}{Victoria Ajayi}, \bibinfo{person}{Victoria Montanez}, \bibinfo{person}{Vijai Mohan}, \bibinfo{person}{Vinay~Satish Kumar},
  \bibinfo{person}{Vishal Mangla}, \bibinfo{person}{Vlad Ionescu}, \bibinfo{person}{Vlad Poenaru}, \bibinfo{person}{Vlad~Tiberiu Mihailescu}, \bibinfo{person}{Vladimir Ivanov}, \bibinfo{person}{Wei Li}, \bibinfo{person}{Wenchen Wang}, \bibinfo{person}{Wenwen Jiang}, \bibinfo{person}{Wes Bouaziz}, \bibinfo{person}{Will Constable}, \bibinfo{person}{Xiaocheng Tang}, \bibinfo{person}{Xiaojian Wu}, \bibinfo{person}{Xiaolan Wang}, \bibinfo{person}{Xilun Wu}, \bibinfo{person}{Xinbo Gao}, \bibinfo{person}{Yaniv Kleinman}, \bibinfo{person}{Yanjun Chen}, \bibinfo{person}{Ye Hu}, \bibinfo{person}{Ye Jia}, \bibinfo{person}{Ye Qi}, \bibinfo{person}{Yenda Li}, \bibinfo{person}{Yilin Zhang}, \bibinfo{person}{Ying Zhang}, \bibinfo{person}{Yossi Adi}, \bibinfo{person}{Youngjin Nam}, \bibinfo{person}{Yu}, \bibinfo{person}{Wang}, \bibinfo{person}{Yu Zhao}, \bibinfo{person}{Yuchen Hao}, \bibinfo{person}{Yundi Qian}, \bibinfo{person}{Yunlu Li}, \bibinfo{person}{Yuzi He}, \bibinfo{person}{Zach Rait}, \bibinfo{person}{Zachary
  DeVito}, \bibinfo{person}{Zef Rosnbrick}, \bibinfo{person}{Zhaoduo Wen}, \bibinfo{person}{Zhenyu Yang}, \bibinfo{person}{Zhiwei Zhao}, {and} \bibinfo{person}{Zhiyu Ma}.} \bibinfo{year}{2024}\natexlab{}.
\newblock \bibinfo{title}{The Llama 3 Herd of Models}.
\newblock
\showeprint[arxiv]{2407.21783}~[cs.AI]
\urldef\tempurl%
\url{https://arxiv.org/abs/2407.21783}
\showURL{%
\tempurl}


\bibitem[Ha et~al\mbox{.}(2024)]%
        {feelingloved}
\bibfield{author}{\bibinfo{person}{Thao Ha}, \bibinfo{person}{Masumi Iida}, \bibinfo{person}{Selena Quiroz}, \bibinfo{person}{Olivia Maras}, {and} \bibinfo{person}{Andrea Savord}.} \bibinfo{year}{2024}\natexlab{}.
\newblock \showarticletitle{Feeling loved in mixed‐gender adolescent romantic relationships in the face of daily stress and support: A dyadic diary study}.
\newblock \bibinfo{journal}{\emph{Developmental Science}}  \bibinfo{volume}{27} (\bibinfo{date}{04} \bibinfo{year}{2024}), \bibinfo{pages}{e13511}.
\newblock
\href{https://doi.org/10.1111/desc.13511}{doi:\nolinkurl{10.1111/desc.13511}}


\bibitem[Huang et~al\mbox{.}(2024)]%
        {emotionRAG}
\bibfield{author}{\bibinfo{person}{Le Huang}, \bibinfo{person}{Hengzhi Lan}, \bibinfo{person}{Zijun Sun}, \bibinfo{person}{Chuan Shi}, {and} \bibinfo{person}{Ting Bai}.} \bibinfo{year}{2024}\natexlab{}.
\newblock \showarticletitle{Emotional RAG: Enhancing Role-Playing Agents through Emotional Retrieval}. In \bibinfo{booktitle}{\emph{2024 IEEE International Conference on Knowledge Graph (ICKG)}}. \bibinfo{pages}{120--127}.
\newblock
\href{https://doi.org/10.1109/ICKG63256.2024.00023}{doi:\nolinkurl{10.1109/ICKG63256.2024.00023}}


\bibitem[Jiang et~al\mbox{.}(2024)]%
        {jiang-etal-2024-personallm}
\bibfield{author}{\bibinfo{person}{Hang Jiang}, \bibinfo{person}{Xiajie Zhang}, \bibinfo{person}{Xubo Cao}, \bibinfo{person}{Cynthia Breazeal}, \bibinfo{person}{Deb Roy}, {and} \bibinfo{person}{Jad Kabbara}.} \bibinfo{year}{2024}\natexlab{}.
\newblock \showarticletitle{{P}ersona{LLM}: Investigating the Ability of Large Language Models to Express Personality Traits}. In \bibinfo{booktitle}{\emph{Findings of the Association for Computational Linguistics: NAACL 2024}}, \bibfield{editor}{\bibinfo{person}{Kevin Duh}, \bibinfo{person}{Helena Gomez}, {and} \bibinfo{person}{Steven Bethard}} (Eds.). \bibinfo{publisher}{Association for Computational Linguistics}, \bibinfo{address}{Mexico City, Mexico}, \bibinfo{pages}{3605--3627}.
\newblock
\href{https://doi.org/10.18653/v1/2024.findings-naacl.229}{doi:\nolinkurl{10.18653/v1/2024.findings-naacl.229}}


\bibitem[Larrea et~al\mbox{.}(2025)]%
        {lovesims}
\bibfield{author}{\bibinfo{person}{Mateo Larrea}, \bibinfo{person}{Xingyi Zhang}, {and} \bibinfo{person}{Xuyang Zhu}.} \bibinfo{year}{2025}\natexlab{}.
\newblock \showarticletitle{LoveSims: Exploring ‘What-If’ Scenarios for Relationship Insights and Compatibility}. In \bibinfo{booktitle}{\emph{Proceedings of the Extended Abstracts of the CHI Conference on Human Factors in Computing Systems}} \emph{(\bibinfo{series}{CHI EA '25})}. \bibinfo{publisher}{Association for Computing Machinery}, \bibinfo{address}{New York, NY, USA}, Article \bibinfo{articleno}{383}, \bibinfo{numpages}{7}~pages.
\newblock
\showISBNx{9798400713958}
\href{https://doi.org/10.1145/3706599.3720011}{doi:\nolinkurl{10.1145/3706599.3720011}}


\bibitem[Li et~al\mbox{.}(2023)]%
        {li2023camel}
\bibfield{author}{\bibinfo{person}{Guohao Li}, \bibinfo{person}{Hasan Abed Al~Kader Hammoud}, \bibinfo{person}{Hani Itani}, \bibinfo{person}{Dmitrii Khizbullin}, {and} \bibinfo{person}{Bernard Ghanem}.} \bibinfo{year}{2023}\natexlab{}.
\newblock \showarticletitle{{CAMEL}: Communicative Agents for ''Mind'' Exploration of Large Language Model Society}. In \bibinfo{booktitle}{\emph{Thirty-seventh Conference on Neural Information Processing Systems}}.
\newblock
\urldef\tempurl%
\url{https://openreview.net/forum?id=3IyL2XWDkG}
\showURL{%
\tempurl}


\bibitem[OpenAI et~al\mbox{.}(2025)]%
        {openai2025gptoss120bgptoss20bmodel}
\bibfield{author}{\bibinfo{person}{OpenAI}, \bibinfo{person}{:}, \bibinfo{person}{Sandhini Agarwal}, \bibinfo{person}{Lama Ahmad}, \bibinfo{person}{Jason Ai}, \bibinfo{person}{Sam Altman}, \bibinfo{person}{Andy Applebaum}, \bibinfo{person}{Edwin Arbus}, \bibinfo{person}{Rahul~K. Arora}, \bibinfo{person}{Yu Bai}, \bibinfo{person}{Bowen Baker}, \bibinfo{person}{Haiming Bao}, \bibinfo{person}{Boaz Barak}, \bibinfo{person}{Ally Bennett}, \bibinfo{person}{Tyler Bertao}, \bibinfo{person}{Nivedita Brett}, \bibinfo{person}{Eugene Brevdo}, \bibinfo{person}{Greg Brockman}, \bibinfo{person}{Sebastien Bubeck}, \bibinfo{person}{Che Chang}, \bibinfo{person}{Kai Chen}, \bibinfo{person}{Mark Chen}, \bibinfo{person}{Enoch Cheung}, \bibinfo{person}{Aidan Clark}, \bibinfo{person}{Dan Cook}, \bibinfo{person}{Marat Dukhan}, \bibinfo{person}{Casey Dvorak}, \bibinfo{person}{Kevin Fives}, \bibinfo{person}{Vlad Fomenko}, \bibinfo{person}{Timur Garipov}, \bibinfo{person}{Kristian Georgiev}, \bibinfo{person}{Mia Glaese},
  \bibinfo{person}{Tarun Gogineni}, \bibinfo{person}{Adam Goucher}, \bibinfo{person}{Lukas Gross}, \bibinfo{person}{Katia~Gil Guzman}, \bibinfo{person}{John Hallman}, \bibinfo{person}{Jackie Hehir}, \bibinfo{person}{Johannes Heidecke}, \bibinfo{person}{Alec Helyar}, \bibinfo{person}{Haitang Hu}, \bibinfo{person}{Romain Huet}, \bibinfo{person}{Jacob Huh}, \bibinfo{person}{Saachi Jain}, \bibinfo{person}{Zach Johnson}, \bibinfo{person}{Chris Koch}, \bibinfo{person}{Irina Kofman}, \bibinfo{person}{Dominik Kundel}, \bibinfo{person}{Jason Kwon}, \bibinfo{person}{Volodymyr Kyrylov}, \bibinfo{person}{Elaine~Ya Le}, \bibinfo{person}{Guillaume Leclerc}, \bibinfo{person}{James~Park Lennon}, \bibinfo{person}{Scott Lessans}, \bibinfo{person}{Mario Lezcano-Casado}, \bibinfo{person}{Yuanzhi Li}, \bibinfo{person}{Zhuohan Li}, \bibinfo{person}{Ji Lin}, \bibinfo{person}{Jordan Liss}, \bibinfo{person}{Lily}, \bibinfo{person}{Liu}, \bibinfo{person}{Jiancheng Liu}, \bibinfo{person}{Kevin Lu}, \bibinfo{person}{Chris Lu},
  \bibinfo{person}{Zoran Martinovic}, \bibinfo{person}{Lindsay McCallum}, \bibinfo{person}{Josh McGrath}, \bibinfo{person}{Scott McKinney}, \bibinfo{person}{Aidan McLaughlin}, \bibinfo{person}{Song Mei}, \bibinfo{person}{Steve Mostovoy}, \bibinfo{person}{Tong Mu}, \bibinfo{person}{Gideon Myles}, \bibinfo{person}{Alexander Neitz}, \bibinfo{person}{Alex Nichol}, \bibinfo{person}{Jakub Pachocki}, \bibinfo{person}{Alex Paino}, \bibinfo{person}{Dana Palmie}, \bibinfo{person}{Ashley Pantuliano}, \bibinfo{person}{Giambattista Parascandolo}, \bibinfo{person}{Jongsoo Park}, \bibinfo{person}{Leher Pathak}, \bibinfo{person}{Carolina Paz}, \bibinfo{person}{Ludovic Peran}, \bibinfo{person}{Dmitry Pimenov}, \bibinfo{person}{Michelle Pokrass}, \bibinfo{person}{Elizabeth Proehl}, \bibinfo{person}{Huida Qiu}, \bibinfo{person}{Gaby Raila}, \bibinfo{person}{Filippo Raso}, \bibinfo{person}{Hongyu Ren}, \bibinfo{person}{Kimmy Richardson}, \bibinfo{person}{David Robinson}, \bibinfo{person}{Bob Rotsted}, \bibinfo{person}{Hadi
  Salman}, \bibinfo{person}{Suvansh Sanjeev}, \bibinfo{person}{Max Schwarzer}, \bibinfo{person}{D. Sculley}, \bibinfo{person}{Harshit Sikchi}, \bibinfo{person}{Kendal Simon}, \bibinfo{person}{Karan Singhal}, \bibinfo{person}{Yang Song}, \bibinfo{person}{Dane Stuckey}, \bibinfo{person}{Zhiqing Sun}, \bibinfo{person}{Philippe Tillet}, \bibinfo{person}{Sam Toizer}, \bibinfo{person}{Foivos Tsimpourlas}, \bibinfo{person}{Nikhil Vyas}, \bibinfo{person}{Eric Wallace}, \bibinfo{person}{Xin Wang}, \bibinfo{person}{Miles Wang}, \bibinfo{person}{Olivia Watkins}, \bibinfo{person}{Kevin Weil}, \bibinfo{person}{Amy Wendling}, \bibinfo{person}{Kevin Whinnery}, \bibinfo{person}{Cedric Whitney}, \bibinfo{person}{Hannah Wong}, \bibinfo{person}{Lin Yang}, \bibinfo{person}{Yu Yang}, \bibinfo{person}{Michihiro Yasunaga}, \bibinfo{person}{Kristen Ying}, \bibinfo{person}{Wojciech Zaremba}, \bibinfo{person}{Wenting Zhan}, \bibinfo{person}{Cyril Zhang}, \bibinfo{person}{Brian Zhang}, \bibinfo{person}{Eddie Zhang}, {and}
  \bibinfo{person}{Shengjia Zhao}.} \bibinfo{year}{2025}\natexlab{}.
\newblock \bibinfo{title}{gpt-oss-120b \& gpt-oss-20b Model Card}.
\newblock
\showeprint[arxiv]{2508.10925}~[cs.CL]
\urldef\tempurl%
\url{https://arxiv.org/abs/2508.10925}
\showURL{%
\tempurl}


\bibitem[Park et~al\mbox{.}(2023)]%
        {DBLP:conf/uist/ParkOCMLB23}
\bibfield{author}{\bibinfo{person}{Joon~Sung Park}, \bibinfo{person}{Joseph~C. O'Brien}, \bibinfo{person}{Carrie~Jun Cai}, \bibinfo{person}{Meredith~Ringel Morris}, \bibinfo{person}{Percy Liang}, {and} \bibinfo{person}{Michael~S. Bernstein}.} \bibinfo{year}{2023}\natexlab{}.
\newblock \showarticletitle{Generative Agents: Interactive Simulacra of Human Behavior}. In \bibinfo{booktitle}{\emph{UIST}}. \bibinfo{pages}{2:1--2:22}.
\newblock
\urldef\tempurl%
\url{https://doi.org/10.1145/3586183.3606763}
\showURL{%
\tempurl}


\bibitem[Roddy et~al\mbox{.}(2020)]%
        {roddy}
\bibfield{author}{\bibinfo{person}{Mckenzie Roddy}, \bibinfo{person}{Lucia Walsh~Pedersen}, \bibinfo{person}{Karen Rothman}, \bibinfo{person}{Gabe Hatch}, {and} \bibinfo{person}{Brian Doss}.} \bibinfo{year}{2020}\natexlab{}.
\newblock \showarticletitle{Meta-Analysis of Couple Therapy: Effects Across Outcomes, Designs, Timeframes, and Other Moderators}.
\newblock \bibinfo{journal}{\emph{Journal of Consulting and Clinical Psychology}}  \bibinfo{volume}{88} (\bibinfo{date}{07} \bibinfo{year}{2020}), \bibinfo{pages}{583}.
\newblock
\href{https://doi.org/10.1037/ccp0000514}{doi:\nolinkurl{10.1037/ccp0000514}}


\bibitem[Shao et~al\mbox{.}(2023)]%
        {shao-etal-2023-character}
\bibfield{author}{\bibinfo{person}{Yunfan Shao}, \bibinfo{person}{Linyang Li}, \bibinfo{person}{Junqi Dai}, {and} \bibinfo{person}{Xipeng Qiu}.} \bibinfo{year}{2023}\natexlab{}.
\newblock \showarticletitle{Character-{LLM}: A Trainable Agent for Role-Playing}. In \bibinfo{booktitle}{\emph{Proceedings of the 2023 Conference on Empirical Methods in Natural Language Processing}}, \bibfield{editor}{\bibinfo{person}{Houda Bouamor}, \bibinfo{person}{Juan Pino}, {and} \bibinfo{person}{Kalika Bali}} (Eds.). \bibinfo{publisher}{Association for Computational Linguistics}, \bibinfo{address}{Singapore}, \bibinfo{pages}{13153--13187}.
\newblock
\urldef\tempurl%
\url{https://aclanthology.org/2023.emnlp-main.814/}
\showURL{%
\tempurl}


\bibitem[Sharabi and CAUGHLIN(2017)]%
        {whatpredictsfirstdate}
\bibfield{author}{\bibinfo{person}{Liesel Sharabi} {and} \bibinfo{person}{JOHN CAUGHLIN}.} \bibinfo{year}{2017}\natexlab{}.
\newblock \showarticletitle{What predicts first date success? A longitudinal study of modality switching in online dating: Modality switching in online dating}.
\newblock \bibinfo{journal}{\emph{Personal Relationships}}  \bibinfo{volume}{24} (\bibinfo{date}{04} \bibinfo{year}{2017}).
\newblock
\href{https://doi.org/10.1111/pere.12188}{doi:\nolinkurl{10.1111/pere.12188}}


\bibitem[Sharabi(2022)]%
        {Sharabi2022FindingLO}
\bibfield{author}{\bibinfo{person}{Liesel~L. Sharabi}.} \bibinfo{year}{2022}\natexlab{}.
\newblock \showarticletitle{Finding Love on a First Data: Matching Algorithms in Online Dating}.
\newblock \bibinfo{journal}{\emph{Harvard Data Science Review}} (\bibinfo{year}{2022}).
\newblock
\urldef\tempurl%
\url{https://api.semanticscholar.org/CorpusID:246377860}
\showURL{%
\tempurl}


\bibitem[Todorov et~al\mbox{.}(2023)]%
        {emotionregulation}
\bibfield{author}{\bibinfo{person}{Emily Todorov}, \bibinfo{person}{Alison Paradis}, {and} \bibinfo{person}{Thao Ha}.} \bibinfo{year}{2023}\natexlab{}.
\newblock \showarticletitle{Emotion Regulation Difficulties and Relationship Satisfaction in Adolescent Couples: The Role of Conflict Resolution Strategies}.
\newblock \bibinfo{journal}{\emph{Journal of Youth and Adolescence}}  \bibinfo{volume}{52} (\bibinfo{date}{05} \bibinfo{year}{2023}), \bibinfo{pages}{1--15}.
\newblock
\href{https://doi.org/10.1007/s10964-023-01787-6}{doi:\nolinkurl{10.1007/s10964-023-01787-6}}


\bibitem[Wang et~al\mbox{.}(2024)]%
        {wang-etal-2024-rolellm}
\bibfield{author}{\bibinfo{person}{Noah Wang}, \bibinfo{person}{Z.y. Peng}, \bibinfo{person}{Haoran Que}, \bibinfo{person}{Jiaheng Liu}, \bibinfo{person}{Wangchunshu Zhou}, \bibinfo{person}{Yuhan Wu}, \bibinfo{person}{Hongcheng Guo}, \bibinfo{person}{Ruitong Gan}, \bibinfo{person}{Zehao Ni}, \bibinfo{person}{Jian Yang}, \bibinfo{person}{Man Zhang}, \bibinfo{person}{Zhaoxiang Zhang}, \bibinfo{person}{Wanli Ouyang}, \bibinfo{person}{Ke Xu}, \bibinfo{person}{Wenhao Huang}, \bibinfo{person}{Jie Fu}, {and} \bibinfo{person}{Junran Peng}.} \bibinfo{year}{2024}\natexlab{}.
\newblock \showarticletitle{{R}ole{LLM}: Benchmarking, Eliciting, and Enhancing Role-Playing Abilities of Large Language Models}. In \bibinfo{booktitle}{\emph{Findings of the Association for Computational Linguistics: ACL 2024}}, \bibfield{editor}{\bibinfo{person}{Lun-Wei Ku}, \bibinfo{person}{Andre Martins}, {and} \bibinfo{person}{Vivek Srikumar}} (Eds.). \bibinfo{publisher}{Association for Computational Linguistics}, \bibinfo{address}{Bangkok, Thailand}, \bibinfo{pages}{14743--14777}.
\newblock
\href{https://doi.org/10.18653/v1/2024.findings-acl.878}{doi:\nolinkurl{10.18653/v1/2024.findings-acl.878}}


\bibitem[Yang et~al\mbox{.}(2025)]%
        {yang2025qwen3technicalreport}
\bibfield{author}{\bibinfo{person}{An Yang}, \bibinfo{person}{Anfeng Li}, \bibinfo{person}{Baosong Yang}, \bibinfo{person}{Beichen Zhang}, \bibinfo{person}{Binyuan Hui}, \bibinfo{person}{Bo Zheng}, \bibinfo{person}{Bowen Yu}, \bibinfo{person}{Chang Gao}, \bibinfo{person}{Chengen Huang}, \bibinfo{person}{Chenxu Lv}, \bibinfo{person}{Chujie Zheng}, \bibinfo{person}{Dayiheng Liu}, \bibinfo{person}{Fan Zhou}, \bibinfo{person}{Fei Huang}, \bibinfo{person}{Feng Hu}, \bibinfo{person}{Hao Ge}, \bibinfo{person}{Haoran Wei}, \bibinfo{person}{Huan Lin}, \bibinfo{person}{Jialong Tang}, \bibinfo{person}{Jian Yang}, \bibinfo{person}{Jianhong Tu}, \bibinfo{person}{Jianwei Zhang}, \bibinfo{person}{Jianxin Yang}, \bibinfo{person}{Jiaxi Yang}, \bibinfo{person}{Jing Zhou}, \bibinfo{person}{Jingren Zhou}, \bibinfo{person}{Junyang Lin}, \bibinfo{person}{Kai Dang}, \bibinfo{person}{Keqin Bao}, \bibinfo{person}{Kexin Yang}, \bibinfo{person}{Le Yu}, \bibinfo{person}{Lianghao Deng}, \bibinfo{person}{Mei Li}, \bibinfo{person}{Mingfeng
  Xue}, \bibinfo{person}{Mingze Li}, \bibinfo{person}{Pei Zhang}, \bibinfo{person}{Peng Wang}, \bibinfo{person}{Qin Zhu}, \bibinfo{person}{Rui Men}, \bibinfo{person}{Ruize Gao}, \bibinfo{person}{Shixuan Liu}, \bibinfo{person}{Shuang Luo}, \bibinfo{person}{Tianhao Li}, \bibinfo{person}{Tianyi Tang}, \bibinfo{person}{Wenbiao Yin}, \bibinfo{person}{Xingzhang Ren}, \bibinfo{person}{Xinyu Wang}, \bibinfo{person}{Xinyu Zhang}, \bibinfo{person}{Xuancheng Ren}, \bibinfo{person}{Yang Fan}, \bibinfo{person}{Yang Su}, \bibinfo{person}{Yichang Zhang}, \bibinfo{person}{Yinger Zhang}, \bibinfo{person}{Yu Wan}, \bibinfo{person}{Yuqiong Liu}, \bibinfo{person}{Zekun Wang}, \bibinfo{person}{Zeyu Cui}, \bibinfo{person}{Zhenru Zhang}, \bibinfo{person}{Zhipeng Zhou}, {and} \bibinfo{person}{Zihan Qiu}.} \bibinfo{year}{2025}\natexlab{}.
\newblock \bibinfo{title}{Qwen3 Technical Report}.
\newblock
\showeprint[arxiv]{2505.09388}~[cs.CL]
\urldef\tempurl%
\url{https://arxiv.org/abs/2505.09388}
\showURL{%
\tempurl}


\bibitem[Zhou et~al\mbox{.}(2024)]%
        {zhou2024sotopia}
\bibfield{author}{\bibinfo{person}{Xuhui Zhou}, \bibinfo{person}{Hao Zhu}, \bibinfo{person}{Leena Mathur}, \bibinfo{person}{Ruohong Zhang}, \bibinfo{person}{Haofei Yu}, \bibinfo{person}{Zhengyang Qi}, \bibinfo{person}{Louis-Philippe Morency}, \bibinfo{person}{Yonatan Bisk}, \bibinfo{person}{Daniel Fried}, \bibinfo{person}{Graham Neubig}, {and} \bibinfo{person}{Maarten Sap}.} \bibinfo{year}{2024}\natexlab{}.
\newblock \showarticletitle{{SOTOPIA}: Interactive Evaluation for Social Intelligence in Language Agents}. In \bibinfo{booktitle}{\emph{The Twelfth International Conference on Learning Representations}}.
\newblock
\urldef\tempurl%
\url{https://openreview.net/forum?id=mM7VurbA4r}
\showURL{%
\tempurl}


\end{thebibliography}

\begin{appendices}
\section{Scene State Dictionary}
\label{appendix:scene_state}
\begin{verbatim}
theme: str
setting: str
NPC: list[str]
current_scene: str
previous_summary: str
character_1_goal: str
character_2_goal: str
scene_conflict: str
\end{verbatim}
\section{Agent Choice from Scene-Managed Options}
\label{appendix:decision_prompt}
\begin{quote}
\textbf{Role.} 
You are currently in a romantic relationship and facing a decision.

\textbf{Instructions:}
\begin{itemize}
\item Carefully review the list of candidate actions. Do not invent or modify options---choose only from those provided.
\item Select the action that best fits the character's personality traits, current motivations, and the central conflict of the scene.
\item Consider the character's recent history and prior events. Ensure your choice maintains narrative continuity and does not contradict what has already happened.
\item Do not include dialogue, internal monologue, or describe future actions by other characters. Focus on a concrete, external action that can be enacted in the next scene.
\end{itemize}

\textbf{Selection criteria (use qualitative judgment, not numeric scoring):}
\begin{itemize}
\item Relevance to the current scene conflict
\item Consistency with the character's personality, goals, and constraints
\item Likelihood to cause a meaningful state change (e.g., in trust, closeness, autonomy, conflict intensity, commitment, resources, or reputation)
\item Plausibility and reversibility within the story context
\end{itemize}
\textbf{Context inputs:}
\begin{verbatim}
Most Recent Internal Thought:
{{ internal_thought }}

Your Persona:
{{ agent_persona }}

Scene History:
{{ current_scene_history }}

Relevant Memories:
{{ top_k_retrieved_memories }}

Action Options:
{{ action_options }}
\end{verbatim}

\textbf{Output:}
\begin{lstlisting}
Output the result as a valid dictionary in the following format.
Do not include any other strings or literals:
{
"action": "realistic, personality-based next action with tone",
"reasoning": "why was this action chosen"
}
\end{lstlisting}
\end{quote}

\section{Emotion Embeddings}
\label{appendix:emotion_embedding}
\begin{figure}[H]
\caption{illustration of emotion embedding pipeline}
\centering
\includegraphics[scale=0.3]{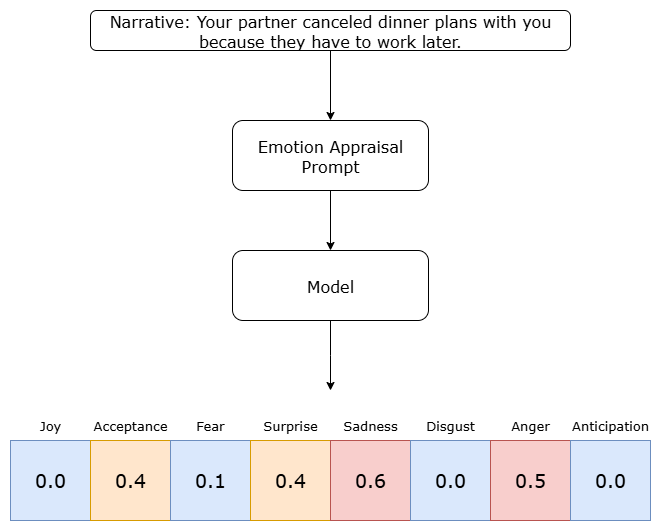} 
\end{figure}

\section{Relationship State Enums}
\label{appendix:state_enums}
\begin{quote}
\texttt{conflict:} "none", "brewing", "active", "unresolved", "repaired", "unknown" \\
\texttt{repair\_outcome:} "none", "attempted", "successful", "failed", "unknown" \\ \texttt{clarity:} "unclear", "tacit", "explicit", "unknown" \\ \texttt{constraints:} "none", "emerging", "accrued", "unknown" \\ \texttt{alternatives:} "quiet", "salient", "hot", "unknown" \\ \texttt{transition:} "none", "upcoming", "underway", "unknown" \\ \texttt{network:} "supportive", "neutral", "opposed", "mixed", "unknown" \\
\texttt{breakup\_marker:} "none", "soft", "hard", unknown''
\end{quote}

\section{Scene Logic}
\label{appendix:scene_logic}
\begin{figure}[H]
\caption{illustration of SceneMaster workflow}
\centering
\includegraphics[scale=0.35]{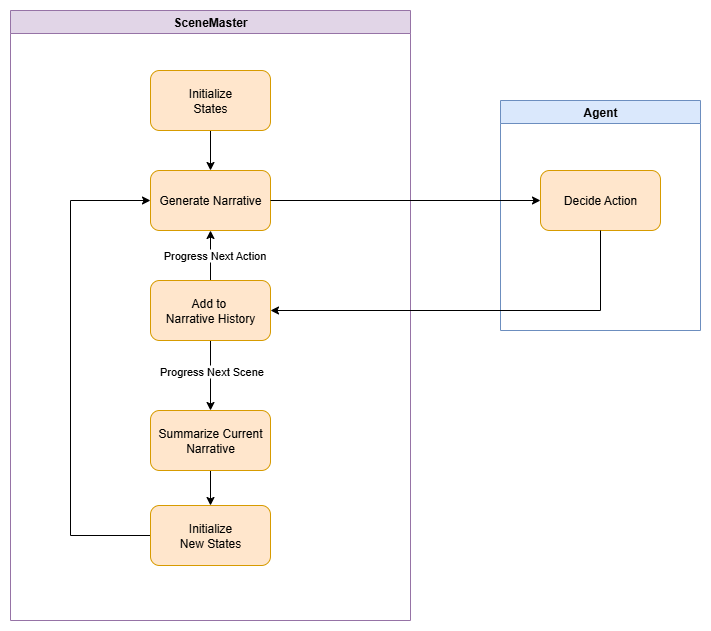} 
\end{figure}

\end{appendices}









\end{document}